\documentclass[%
 reprint,
]{revtex4-2}
\usepackage{amsmath, amsthm, amssymb}
\usepackage{adjustbox}
\usepackage{float,epsfig}
\usepackage{graphicx}
\usepackage{yfonts}
\usepackage{bbm}
\usepackage{rotating}
\newcommand{\mycommand}[1]{\csname #1command\endcsname}
\usepackage{subfigure}
\usepackage{appendix}
\usepackage{caption}
\usepackage{comment}
\usepackage{enumitem}
\usepackage{color}
\usepackage{algorithm}
\usepackage[utf8]{inputenc}
\usepackage{braket}
\usepackage{qcircuit}
\usepackage[overload]{empheq}
\usepackage{hyperref}
\usepackage{xcolor}
\usepackage[english]{babel}
\usepackage[flushleft]{threeparttable}
\usepackage[normalem]{ulem}

\usepackage{algcompatible}
\usepackage[noend]{algpseudocode}
\usepackage{mathtools}

\usepackage{xparse}

\NewDocumentCommand{\INTERVALINNARDS}{ m m }{
    #1 {,} #2
}
\NewDocumentCommand{\interval}{ s m >{\SplitArgument{1}{,}}m m o }{
    \IfBooleanTF{#1}{
        \left#2 \INTERVALINNARDS #3 \right#4
    }{
        \IfValueTF{#5}{
            #5{#2} \INTERVALINNARDS #3 #5{#4}
        }{
            #2 \INTERVALINNARDS #3 #4
        }
    }
}

\usepackage{lipsum}

\usepackage{qcircuit}



\begin{document}
\newtheorem{theorem}{\bf Theorem}[section]
\newtheorem{proposition}[theorem]{\bf Proposition}
\newtheorem{definition}[theorem]{\bf Definition}
\newtheorem{corollary}[theorem]{\bf Corollary}
\newtheorem{example}[theorem]{\bf Example}
\newtheorem{exam}[theorem]{\bf Example}
\newtheorem{remark}[theorem]{\bf Remark}
\newtheorem{lemma}[theorem]{\bf Lemma}
\newtheorem{statement}[theorem]{\bf Statement}
\newcommand{\nrm}[1]{|\!|\!| {#1} |\!|\!|}

\newcommand{\calL}{{\mathcal L}}
\newcommand{\calX}{{\mathcal X}}
\newcommand{\calA}{{\mathcal A}}
\newcommand{\calB}{{\mathcal B}}
\newcommand{\calC}{{\mathcal C}}
\newcommand{\calK}{{\mathcal K}}
\newcommand{\C}{{\mathbb C}}
\newcommand{\R}{{\mathbb R}}
\newcommand{\U}{{\mathrm U}}
\renewcommand{\SS}{{\mathbb S}}
\newcommand{\LL}{{\mathbb L}}
\newcommand{\st}{{\star}}
\def\kernel{\mathop{\rm kernel}\nolimits}
\def\sigan{\mathop{\rm span}\nolimits}

\newcommand{\klasse}{{\boldsymbol \Delta}}

\newcommand{\ba}{\begin{array}}
\newcommand{\ea}{\end{array}}
\newcommand{\von}{\vskip 1ex}
\newcommand{\vone}{\vskip 2ex}
\newcommand{\vtwo}{\vskip 4ex}
\newcommand{\dm}[1]{ {\displaystyle{#1} } }

\newcommand{\be}{\begin{equation}}
\newcommand{\ee}{\end{equation}}
\newcommand{\beano}{\begin{eqnarray*}}
\newcommand{\eeano}{\end{eqnarray*}}
\newcommand{\inp}[2]{\langle {#1} ,\,{#2} \rangle}
\def\bmatrix#1{\left[ \begin{matrix} #1 \end{matrix} \right]}
\def \noin{\noindent}
\newcommand{\evenindex}{\Pi_e}

\newcommand{\tb}[1]{\textcolor{blue}{ #1}}
\newcommand{\tm}[1]{\textcolor{magenta}{ #1}}
\newcommand{\tre}[1]{\textcolor{red}{ #1}}
\newcommand{\snote}[1]{\textcolor{blue}{Shantanav: #1}}



\def \K{{\mathbf k}}
\def \N{{\mathbb N}}
\def \R{{\mathbb R}}
\def \F{{\mathbb F}}
\def \C{{\mathbb C}}
\def \Q{{\mathbb Q}}
\def \Z{{\mathbb Z}}
\def \I{{\mathbb I}}
\def \D{{\mathcal D}}
\def \H{{\mathcal H}}
\def \P{{\mathcal P}}
\def \M{{\mathcal M}}
\def \B{{\mathcal B}}
\def \O{{\mathcal O}}
\def \calG{{\mathcal G}}
\def \PO{{\mathcal {PO}}}
\def \X{{\mathcal X}}
\def \Y{{\mathcal Y}}
\def \calW{{\mathcal W}}
\def \pf{{\bf Proof: }}
\def \lam{{\lambda}}
\def\lc{\left\lceil}   
\def\rc{\right\rceil}
\def \N{{\mathbb N}}
\def \Ls{{\Lambda}_{m-1}}
\def \Gb{\mathrm{G}}
\def \Hb{\mathrm{H}}
\def \Delta{\triangle}
\def \Rar{\Rightarrow}
\def \p{{\mathsf{p}(\lam; v)}}

\def \D{{\mathbb D}}

\def \tr{\mathrm{Tr}}
\def \cond{\mathrm{cond}}
\def \lam{\lambda}
\def \sig{\sigma}
\def \sign{\mathrm{sign}}

\def \ep{\epsilon}
\def \diag{\mathrm{diag}}
\def \rev{\mathrm{rev}}
\def \vec{\mathrm{vec}}

\def \ham{\mathsf{Ham}}
\def \herm{\mathsf{Herm}}
\def \sym{\mathsf{sym}}
\def \odd{\mathsf{sym}}
\def \en{\mathrm{even}}
\def \rank{\mathrm{rank}}
\def \pf{{\bf Proof: }}
\def \dist{\mathrm{dist}}
\def \rar{\rightarrow}

\def \rank{\mathrm{rank}}
\def \pf{{\bf Proof: }}
\def \dist{\mathrm{dist}}
\def \Re{\mathsf{Re}}
\def \Im{\mathsf{Im}}
\def \re{\mathsf{re}}
\def \im{\mathsf{im}}

\def \sym{\mathsf{sym}}
\def \sksym{\mathsf{skew\mbox{-}sym}}
\def \odd{\mathrm{odd}}
\def \even{\mathrm{even}}
\def \herm{\mathsf{Herm}}
\def \skherm{\mathsf{skew\mbox{-}Herm}}
\def \str{\mathrm{ Struct}}
\def \eproof{$\blacksquare$}

\def \cnot{\mathrm{CNOT}}
\definecolor{darkolivegreen}{rgb}{0.33, 0.42, 0.18}

\def \bS{{\bf S}}
\def \cA{{\cal A}}
\def \E{{\mathcal E}}
\def \X{{\mathcal X}}
\def \F{{\mathcal F}}
\def \cH{\mathcal{H}}
\def \cJ{\mathcal{J}}
\def \tr{\mathrm{Tr}}
\def \range{\mathrm{Range}}
\def \adj{\star}

\pdfstringdefDisableCommands{%
  \def\\{}%
  \def\texttt#1{<#1>}%
}
\def \adj{\star}

\def \pal{\mathrm{palindromic}}
\def \palpen{\mathrm{palindromic~~ pencil}}
\def \palpoly{\mathrm{palindromic~~ polynomial}}
\def \odd{\mathrm{odd}}
\def \even{\mathrm{even}}

\newcommand{\tg}[1]{\textcolor{green}{ #1}}

\preprint{APS/123-QED}

\title{A Fully Device-Independent Ternary Quantum Key Distribution Protocol Based on the Impossible Colouring Game}

\author{Aniket Basak$^{1}$}
\email{$^1$ aniket2001basak@gmail.com}
\author{Rajeet Ghosh$^{2}$}
\email{$^2$ rajeet.ghosh@students.iiserpune.ac.in}
\author{Rohit Sarma Sarkar$^3$}
\email{$^4$ rohit15sarkar@yahoo.com(Corresponding Author)}
\author{Chandan Goswami$^{4}$}
\email{$^3$ 7cgoswami@gmail.com}
\author{Avishek Adhikari$^5$}
\email{$^5$ avishek.adh@gmail.com}

\affiliation{$^1$Cryptology and Security Research Unit, Indian Statistical Institute Kolkata, Kolkata 700108, India}
\affiliation{$^2$Department of Mathematics, Indian Institute of Technology Madras, Chennai 600036, India}
\affiliation{$^3$
Universitat Politècnica de Catalunya, Barcelona 08034, Spain}
\affiliation{$^{4}$Department of Mathematics, Indian Institute of Technology Kharagpur, Kharagpur 721302, India}
\affiliation{$^{5}$Department of Mathematics, Presidency University, Kolkata 700073, India}



\begin{abstract}
We propose a Ternary Fully Device-Independent Quantum Key Distribution (TFDIQKD) protocol based on the two-party Impossible Colouring pseudo-telepathy game, utilizing maximally entangled qutrit states to enable secure key generation between distant parties. The protocol harnesses Bell inequality violations that arise from contextuality in the Kochen-Specker theorem, thereby offering a quantum advantage in a task that is classically impossible and eliminating reliance on assumptions about the internal functioning of quantum devices. A specially designed qutrit quantum circuit is used for state preparation. Security and device independence are rigorously analyzed within a composable framework, employing Bell-inequality violations, smooth min-entropy, von Neumann entropy, and Shannon entropy. The protocol achieves optimal key rates in the ideal case and maintains security under significant noise, with a finite-key analysis that supports its practical viability. Overall, the protocol operates within an adequate security framework and demonstrates an improved key generation rate compared to standard quantum key distribution schemes, highlighting the potential of high-dimensional quantum systems for secure communication.

\end{abstract}

\maketitle
\thispagestyle{empty}

\section{Introduction}\label{introduction}

 Quantum Key Distribution (QKD) \cite{AGM06, BB84, JBS19} is a secure communication protocol utilising the fundamental principles of quantum mechanics.  It has emerged as a crucial solution, enabling two remote parties to establish a shared, secret key with {information-theoretic} security. In other words, this guarantees unconditional security based on the fundamental laws of quantum mechanics, making cryptographic protocols immune to attacks regardless of an adversary’s computational power. The need for QKD stems from the vulnerability that quantum computers pose to classical cryptographic systems. While some symmetric key ciphers remain resistant to quantum attacks, particularly against Grover’s search~\cite{GROVER96} and counting algorithms~\cite{NC10}, the traditional approach of using public-key cryptography to share symmetric keys is now vulnerable. In order to address this challenge, two approaches have surfaced in literature. The first involves designing classical post-quantum cryptographic schemes based on quantum-resistant hardness assumptions. While these systems offer strong security, they often introduce computational overhead, particularly in signature schemes and bandwidth usage, though they remain practical for real-world deployment. 
 The second approach is to utilize the inherent hardness of quantum mechanics in order to create an unconditionally secure key for communication over public channels and to protect against both classical and quantum attacks. This quantum approach primarily relies on two fundamental principles, viz. the security provided by quantum measurement and the security derived from quantum entanglement. 

The security of QKD relies on the no-cloning theorem~\cite{WZ09} and quantum measurement-induced disturbance, ensuring that any eavesdropping attempt alters the quantum state, making interception detectable. QKD offers the advantage of generating fresh, secure cryptographic keys that can be immediately utilized. In order to achieve this, a significant final key must be generated in a very short time frame. A significant benefit of QKD is that, since communication is quantum, an adversary does not retain a classical transcript once a QKD session concludes. The fundamental steps of QKD protocols include quantum state distribution, parameter estimation, block-wise measurement, and classical postprocessing. The concept of QKD was first introduced by Bennett and Brassard in 1984 with their BB84 protocol~\cite{BB84}. Since then, numerous modifications, extensions, and variants have been proposed~\cite{AGM06, AAM06, BC92, TMF13,TMS13, TMCNR12}. In 1991, Ekert~\cite{EA91} introduced the first entanglement-based QKD protocol. The security of these protocols fundamentally relies on the assumption that the communicating parties use either single-photon sources or maximally entangled states. However, conventional QKD protocols, such as BB84, face challenges in practical implementation due to factors like noise and limited key rates.  To overcome these limitations, Device-Independent Quantum Key Distribution (DIQKD) has gained prominence, offering security guarantees without assumptions about the internal workings of quantum devices. DIQKD protocols typically rely on Bell non-locality tests, but their efficiency is often constrained by low key rates. 

In DIQKD\cite{MS16, ZM23}, two honest parties at remote locations aim to establish a secure cryptographic key. Unlike traditional key distribution protocols, the security of a QKD is based on the fundamental principles of quantum mechanics. For DIQKD, in particular, the emphasis is on reducing trust in the internal working of the devices utilized by the communication parties. Rather than assuming optimal implementation, DIQKD protocols derive security directly from observable statistics, frequently by testing for non-classical correlations that signal the presence of quantum entanglement or system dimension limitations. This design allows key distribution even when devices are imperfect or partially untrusted. In this paradigm, all devices involved in preparing, transmitting, and measuring information carriers are treated as black boxes, meaning they could have been created or manipulated by an adversary. This ensures security even when the internal workings of the devices are unknown or potentially compromised. It is to be noted that, despite its name, DIQKD can still assume some level of trust in certain components. For example, Measurement-Device-Independent QKD (MDI-QKD) assumes trusted state preparation but removes trust from the measurement devices. The fundamental requirement is that the two honest parties remain spatially separated to prevent hidden communication between their devices. Other works like memory-assisted MDI-QKD protocols \cite{PCR14} in literature.

There also exist versions of DIQKD, known as fully DIQKD (FDIQKD) \cite{VV19}, where one eliminates any trust in all quantum devices, including the source and preparation of quantum states, which could potentially be adversarially controlled. In this paradigm, the only assumption on the devices is that they can be modeled by the laws of quantum mechanics, and that they are spatially isolated from each other and from any adversary's laboratory. Also, the devices may have quantum memory \cite{VV14}. FDIQKD ensures security purely from the violation of Bell inequalities \cite{bell14,JS16}. It is considered to be the strongest form of device-independent key distribution. While  DIQKD in general and FDIQKD in particular rely on spatial separation and quantum correlations, the latter provides stronger security guarantees by removing even minimal trust in state preparation through the use of Bell inequalities.

However, this increased security comes with greater experimental challenges, making FDIQKD more difficult to implement in practice\cite{VV14, VV19}. FDIQKD requires sources of quantum states that are completely untrusted, meaning security must be established purely through Bell inequality violations. This is extremely challenging because state preparation is often imperfect in real-world implementations\cite{VV14, VV19}. In contrast, protocols that relax these assumptions, such as those allowing partial trust in device components or those using dimension-based constraints, offer more practical implementations but at the cost of reduced security. These variants serve as intermediate steps but do not achieve the same level of device independence as FDIQKD. 

Moreover, FDIQKD is not constrained by dimension assumptions or detailed device models, ensuring that its security proofs approach the theoretical optimum dictated by quantum mechanics\cite{VV14,VV19}. By closing all detection-related loopholes, it offers foolproof security against increasingly sophisticated attacks\cite{PPG23,ZLR22}. While current technology still faces hurdles in terms of detection efficiency and channel losses, rapid advances in quantum hardware are steadily reducing these limitations. Thus, FDIQKD stands as the most secure paradigm available today, and continued progress in experimental platforms is expected to make it increasingly practical for real-world deployment\cite{ZLA23}.


Thus, in this work, we propose a Ternary FDIQKD (TFDIQKD) protocol based on the Impossible Colouring pseudo-telepathy game~\cite{BBT05}. By employing an appropriate quantum strategy, both parties can achieve a unit probability of winning, surpassing the success rate of other non-local games such as the CHSH game~\cite{CHSH69}. In this {theoretical framework}, we introduce a novel approach to enhancing both security and efficiency in quantum cryptography. {Unlike conventional qubit-based QKD schemes, our protocol operates on qutrits (three-level quantum systems),} which offer sending of larger information for a given transmission in a channel {\cite{SS10}}, improved noise resilience {\cite{BT00,CBKA02}}, and stronger security \cite{SS10} due to more complex quantum correlations. Also, the implementation is feasible through the use of angular momentum modes \cite{MG04} or time bins \cite{TAZG04}\cite{ILC19}. They have also been generated using biphotons, as shown by Bogdanov et al. \cite{BCKM04}. The Impossible Colouring Game serves as the foundation of our protocol. This provides a unit quantum-winning probability while being classically unwinnable, which points to an inherently secure key exchange. There already exist several DIQKD protocols using qutrits \cite{BHA00,KDMK03,DCGZ03,JS16}. {However, our proposed Ternary Fully Device-Independent Quantum Key Distribution is a novel implementation of QKD using a qutrit-based pseudo-telepathy game}, significantly advancing the security and efficiency of quantum cryptographic systems. We also establish the security of our protocol through a rigorous analysis based on smooth minimum entropy, von Neumann entropy, and Shannon entropy, demonstrating its robustness against quantum attacks. 
It is of note that, unlike the binary system of classical computers, superconducting qubits \cite{yan19} and trapped-ion quantum computers \cite{Ald20} theoretically possess discrete energy levels spanning an infinite spectrum, making them naturally suited for qudit-based computations. Further, qutrits (or qudits), contrary to their classical counterpart known as the ternary bits (or trits), can be realistically generated and store more information than qubits. While numerous quantum walk experiments have been conducted on real quantum hardware using qubits \cite{balu18,yan19, Ald20,acasiete20,singh20}, efficiently implementing similar experiments with qutrits and higher-dimensional qudits remains a significant challenge \cite{zhou19}. Nonetheless, qutrit-based circuit models have garnered substantial interest as an alternative to qubit circuits due to their potential for resource-efficient computation \cite{gokhale19,gokhale20}. Recent advancements in hybrid quantum circuits suggest that incorporating intermediate qutrits instead of qubits enables more efficient decomposition of $n$-qubit unitary gates \cite{gokhale19,majumdar22}. Notably, qutrit-based synthesis has demonstrated logarithmic depth reductions, leading to an exponential decrease in resource requirements. Specifically, qutrit circuits require 70 times fewer two-qudit gates compared to conventional two-qubit (CNOT) gates. Furthermore, qutrit-based processors have been theoretically shown to enhance quantum error correction with smaller code sizes \cite{murali17,campbell14} and play a pivotal role in high-fidelity magic state distillation \cite{campbell12}, robust quantum cryptography \cite{bechmann00,bruss02}, and secure quantum communication protocols \cite{vaziri02}. Recent benchmarking efforts on a five-qutrit processor have reported remarkably low single-qutrit gate infidelity \cite{morvan21}, indicating promising advancements toward practical qutrit-based computation. Literature also exists on quantum walk circuit constructions \cite{douglas09}, including qutrit circuits designed for three-state lazy quantum walks on a line \cite{saha21} and qudit circuits for $d$-state discrete-time quantum walks (DTQWs) on a lattice \cite{saha24}. Further, the generalization of quantum algorithms—such as Shor’s algorithm and the Deutsch-Josza algorithm—to qutrit systems has been explored \cite{fan07,bocharov16,bocharov17}. A particularly notable development is the recent formulation of the Quantum Approximate Optimization Algorithm (QAOA) for solving the graph 3-coloring problem using qutrits, which significantly reduces circuit depth and entangling gate count per layer \cite{bottrill23}. Moreover, methods for synthesizing qudit gates have been detailed in \cite{di13}, further enriching the foundation for scalable quantum computing using qutrits and higher-dimensional qudits. In~\cite{sarkar2020periodicity}, the authors also provided a scalable qutrit circuit model for simulating discrete-time quantum walks on Cayley graphs of dihedral groups. Qutrits also offer superior security compared to qubit-based QKD, as their higher-dimensional quantum correlations enhance resilience against adversarial attacks. 

Despite the current challenges in constructing qutrit-based QKD systems, primarily due to technological constraints, our approach establishes a strong theoretical foundation for future implementations. A fundamental challenge that is present is the lack of well-established simulation frameworks for qutrit-based systems, making practical verification uncertain at present. Although platforms like Google Cirq~\cite{kushnarev2025qudit} allow for qutrit computations, a lot of gates have to be defined manually at the start of the program, making the endeavour quite cumbersome. However, recent developments of Google’s Cirq framework supporting qudit-based computations suggest that the feasibility of qutrit-based QKD simulations may soon become a reality, opening new avenues for experimental validation.

One of the key advantages of QKD is the ability to generate fresh, secure keys in real-time, allowing immediate utilization for secure communication. Achieving this requires a significant final key generation rate within a matter of seconds. Our protocol achieves the highest key rate among existing DIQKD protocols 
while maintaining optimal bit wastage and significantly improved noise tolerance. Specifically, we show that our protocol achieves a higher key rate (see Section \ref{IV.d}) than even the BB84 protocol \cite{BB84}, particularly in the ideal case of zero noise. Subsequently, we generalize Renner et. al's \cite{RR08} qubit-based key rate formula for one-way QKD protocols, extending its applicability to qutrit systems. The proposed protocol works on
a maximally entangled qutrit state, enabling more efficient information encoding, increased computational complexity, and greater robustness\cite{DCGZ03} against any variants of noise. Since employing higher-dimensional systems enhances the security of QKD protocols by introducing more complex correlations that are more difficult for an eavesdropper to exploit, the qutrit-based protocol also helps in enhancing the security of QKD protocols by generating more intricate correlations that are difficult for an eavesdropper to intercept, while simultaneously enabling higher key rates. 

We also demonstrate that our protocol is $\epsilon$-correct (Definition \ref{def1}). If an eavesdropper has a strong correlation with the system shared by the two parties, this can be detected during the testing phase. The successful completion of this phase ensures that there is no (or negligible) correlation, thus affirming the DIQKD protocol's efficacy. Moreover, the output of our proposed protocol, i.e., the raw key, is at most $\eta$-distinguishable {(Definition \ref{eta})}. We have also validated the security of our proposed scheme, utilizing the special case of the Impossible Colouring game described in section \ref{sec2} through the state-of-the-art security framework proposed by Renner \cite{RR08, RGK05} and also used by  Xu et al.~\cite{XMZ20} and Tomamichel et al.~\cite{TL17}. In fact, the security analysis of our protocol follows the standard theoretical proof, which is based on the concepts of smooth minimum-entropy, von Neumann entropy, and binary Shannon entropy \cite{NC10}. Furthermore, we have also carried out a finite-key analysis for the same.

Our proposed protocol also outperforms existing DIQKD protocols and exhibits significantly high noise tolerance. Moreover, we also establish that our key rate remains independent of the number of input subsystems, making it constant regardless of the choice of input subsystems. For a comparative performance analysis of our protocol with other state-of-the-art DIQKD protocols, see Table \ref{tab3.5}. 

The paper is organized as follows. In Section \ref{sec2}, we introduce the basic structure of generic device-independent quantum key distribution protocols, followed by a discussion of the two-party impossible coloring game in Section \ref{2A}. In Section \ref{sec2.1}, we present our proposed device-independent quantum key distribution protocol. An illustrative example is provided in Section \ref{exam}, and the complete protocol, along with the raw key generation procedure using maximally entangled states, is detailed in Section \ref{3B}. Section \ref{secqtr} describes the quantum circuit construction required for preparing maximally entangled qutrit Bell states. Section \ref{sec4} presents the security analysis, including correctness, device independence, and the security of both the raw and active keys. The paper concludes in Section \ref{conclusion}.


\section{Preliminaries}\label{sec2}


 In this section, we shall provide background on the two-party impossible colouring game, the generic DIQKD protocol, and other fundamental aspects of the protocol that will be used in the later part of the paper.

Firstly, we shall take a look at the history of DIQKD protocols. In 2014, Vazirani et al.\cite{VV14} introduced a device-independent quantum Key Distribution (DIQKD) protocol based on the Clauser-Horne-Shimony-Holt (CHSH) game\cite{CHSH69} to certify the maximality of entanglement. Later, in 2019, Basak et al.\cite{BMM19} and, in 2023, {Zhen et al.\cite{ZM23} proposed two DIQKD protocols utilizing the Parity game \cite{BBT05} and the Mermin-Peres Magic Square game \cite{MD90}, respectively. Both of these games have a unit probability of being won using a quantum strategy (compared to 0.85 in CHSH) and their protocols have been shown to be secure as well.} Furthermore, these protocols allow key generation using a single measurement basis, unlike the  VVQKD (Vazirani-Vidick Quantum Key Distribution) protocol\cite{VV14, VV19}. However, the key rate in these DIQKD protocols remains low because it depends on the number of subsystems involved in the measurement process. When fewer subsystems are used, the key rate decreases. Further, the games used in these protocols can be won with some probability using classical strategies, making them susceptible to classical attacks. To address these limitations, we propose a DIQKD protocol based on the Impossible Colouring game, which offers improved security and key generation efficiency.

\begin{figure}[H]
\centering
\includegraphics[height= 7cm, width= 7.5cm]{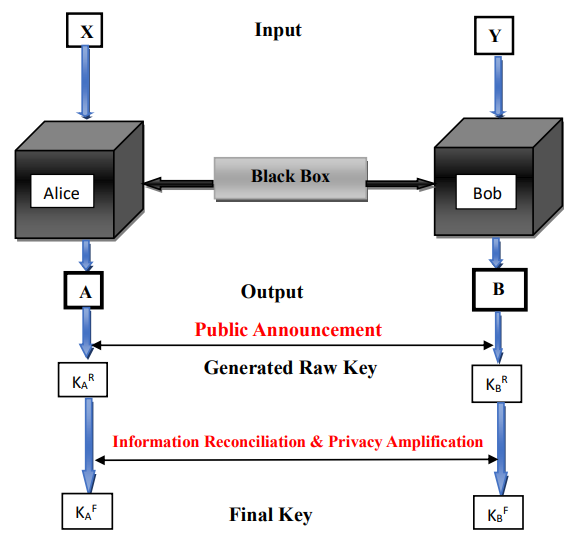}
\caption{Generic DIQKD Protocol.}\label{Fi}
\end{figure}

In Figure~\ref{Fi}, following the footsteps of \cite{PR22, RR08}, we depict a schematic diagram representation of a generic DIQKD protocol.

The road-map in Figure\ref{Fi} depicts that after raw Key Distribution, the process must proceed through parameter estimation, blockwise measurement, information reconciliation, and privacy amplification to generate the final shared secret key securely. We recall that parameter estimation protocol is a two-party procedure in which Alice and Bob, upon receiving inputs from respective Hilbert spaces, perform measurements or computations to estimate certain parameters of the shared system, and based on the estimated values, either output accept or abort the protocol. Information reconciliation is the process by which two parties with similar but imperfectly correlated data communicate in order to correct discrepancies and agree on a common value without revealing the original data. In this context, Alice retains her input $x$, while Bob, holding a correlated input, uses communication to produce an output $x'$ that matches Alice’s value as closely as possible. Privacy amplification is a fundamental technique for extracting a secure key from a partially secret string. It involves applying a randomly chosen hash function from a two-universal family to the input. The resulting key is provably secure, provided that its length is less than the adversary’s uncertainty about the input, as quantified by the (smooth) minimum-entropy.

The Impossible Colouring game \cite{BBT05} is a special type of non-local game in which a quantum strategy enables two players to win with unit probability, while winning with certainty is impossible using classical strategies. Typically, two communicating parties perform a non-locality test~\cite{BCP14} to assess the extent of an adversary’s potential knowledge about the generated data. The results of this test are then used to decide whether the data is suitable for generating secure keys \cite{AR18, MS16}. A detailed description of the two-party Impossible Colouring game is provided in the next section.



\subsection{Two Party Impossible Colouring Game}\label{2A} 

In this section, we describe the Impossible Colouring game, which forms the central idea behind our proposed DIQKD protocol. In order to understand the Impossible Colouring game~\cite{BBT05}, it is necessary to first be familiar with the Kochen-Specker theorem~\cite{peres97}.

\begin{theorem}[\cite{peres97}]\label{kochen}
\label{theorem}
    In $\mathbb{R}^d (d\ge3)$, there is a finite and explicit set of vectors that cannot be $\{0,1\}$-coloured, such that the two following criteria hold simultaneously.
\begin{enumerate}
    \item[i.]  For every orthogonal pair of vectors, at most one is coloured $1$
    \item[ii.] For every mutually orthogonal triple of vectors, at least one of them and therefore exactly one is coloured $1$.
\end{enumerate}
\end{theorem}

In order to explain Theorem \ref{kochen} and subsequently establish our results, we consider the following. Let $V$ denote the subset of vectors in $\mathbb{R}^d$ ($d \geq 3$) that satisfy the conditions of the Kochen-Specker theorem (Theorem \ref{kochen}). 


In the Impossible Colouring game \cite{BBT05}, Alice receives a mutually orthogonal $d$-tuple of vectors $v_1, v_2, \ldots, v_d$, where all vectors are elements of $V$. Bob, on the other hand, receives a single vector $v_\ell$, which is also taken from $V$. The promise of the game is that $v_\ell$ is always one of the vectors given to Alice. They color the vectors as $\{0,1\}$ which fulfills the following criteria:

\begin{enumerate}
    \item[i.] Any two orthogonal vectors in $V$ are not both coloured $1$
    \item[ii.] Exactly one of Alice’s vectors is coloured $1$ if she is given $d$ vectors
    \item[iii.] Alice and Bob assign the same colour to the vector
    $v_\ell$.
\end{enumerate}
In order to meet the aforementioned requirements, if Alice is given $d$ vectors, she should produce $a \in \{1, 2, \ldots, d\}$ as her output, indicating that color $1$ should be assigned to the vector $v_a$. Similarly, Bob, upon receiving his input, outputs a single bit $b$ corresponding to the color assigned to the vector $v_\ell$. To win the game, it is necessary that both Alice and Bob assign the same color to $v_\ell$.

Here, no communication is allowed among the participants after receiving their respective inputs and before producing their outputs. It has been proved in \cite{BBT05} that there does not exist any classical winning strategy for this game. However, quantum entanglement provides an advantage as it has been shown that there exists a perfect quantum strategy by which both parties can always win the game\cite{BBT05}.

 While describing the quantum winning strategy for this game, let us define the necessary shared entangled state as \begin{equation}\label{enst}
     \ket{\psi}= \frac{1}{\sqrt{d}}\sum\limits_{j=0}^{d-1} \ket{j}\ket{j}
 \end{equation}%


Both parties are allowed to share the aforementioned entangled state (Equation \ref{enst}) prior to the start of the game. Upon receiving her input, Alice is given two orthogonal vectors $v_1$ and $v_2$ from the set $V$. She then selects $d-2$ additional vectors, denoted by $v_3, v_4, \ldots, v_d$, which are not necessarily elements of $V$, such that the set $\{v_1, v_2, \ldots, v_d\}$ forms an orthogonal $d$-tuple. Subsequently, she performs a measurement on her share of the entangled state with respect to the basis $B_a$, which consists of these $d$ vectors after proper orthonormalization, if necessary. Similarly, Bob receives a single vector $v_\ell$ and selects $d-1$ additional vectors, again not necessarily from $V$, to complete an orthogonal $d$-tuple. He then performs a measurement on his share of the entangled state with respect to the corresponding orthonormal basis.

After performing their respective measurements, we consider the probability that Alice obtains the outcome corresponding to $v_i$ ($i \in \{1,2,\ldots,d\}$) while Bob obtains the outcome corresponding to $v_\ell$. By following suitable mathematical computations as outlined in \cite{BBT05}, it can be shown that Bob measures $v_\ell$ if and only if Alice also measures $v_\ell$. Consequently, Alice and Bob assign the same color to the vector they have received in common, allowing them to win the game with certainty, without any communication between them.
For our proposed protocol, we consider the pseudo-telepathy game with $d=3$, where $V = V^{\text{KS}}$ is a subset of $\mathbb{R}^3$. Originally, Theorem \ref{theorem} was proved using $117$ vectors \cite{KS90}. Few modifications were done by reducing it to 37 vectors \cite{JB99}, 33 vectors \cite{jb96,KP95}. Later, Conway and Kochen reduced this to $31$ vectors grouped into $17$ orthogonal triples \cite{peres97} which we have used in our construction.

In this work, we adhere to the fixed set of notations depicted in Table \ref{tab:notations}. Any additional symbols will be introduced and explained as they appear in the relevant sections.

\section{Description of Proposed TFDIQKD Protocol}
\label{sec2.1}
Alice and Bob aim to communicate over a potentially insecure channel to which an eavesdropper, Eve, may have access. To securely establish a shared secret key over such an unreliable or insecure channel, they require a QKD protocol. Ideally, the key should be generated without relying on trusted devices. Furthermore, they seek to generate a shared secret key that is classically impossible to compromise and achieves a sufficiently high key rate. These considerations motivate the development of the DIQKD protocol proposed in this work.

In our proposed DIQKD protocol, we assume that Alice and Bob each possess three black boxes, denoted by $D_{1}^{A}$, $D_{2}^{A}$, $D_{3}^{A}$ for Alice and $D_{1}^{B}$, $D_{2}^{B}$, $D_{3}^{B}$ for Bob, respectively. The first two black boxes of Alice accept inputs $\ket{v_1}$ and $\ket{v_2}$, where $v_1, v_2 \in V^{\text{KS}}$ (a subset of $\mathbb{R}^3$ satisfying the Kochen-Specker property), such that $\ket{v_1}$ and $\ket{v_2}$ are orthogonal. The remaining black box of Alice accepts as input a vector $\ket{v_a}$, where $v_a$ is any vector in $\mathbb{R}^3$ (not necessarily from $V^{\text{KS}}$), such that $\ket{v_a}$ is orthogonal to both $\ket{v_1}$ and $\ket{v_2}$.

Alice sends the first two vectors selected from $V^{\text{KS}}$ to Bob one by one through a public channel, which is assumed to be accessible to Eve. Before starting the game, Alice informs Bob which of the received vectors corresponds to $\ket{v_1}$ and which corresponds to $\ket{v_2}$, which ensures that Bob selects his measurement basis correctly.

Subsequently, Bob randomly selects one of the vectors received from Alice as the input for his first black-box ($D_{1}^{B}$), denoted by $\ket{v_\ell}$. The other two black boxes accept inputs $\ket{v_{b_{1}}}$ and $\ket{v_{b_{2}}}$, where $v_{b_{1}}$ and $v_{b_{2}}$ are vectors in $\mathbb{R}^3$ (not necessarily elements of $V^{\text{KS}}$) that are orthogonal to $v_\ell$. We assume that the devices, as in previous device-independent QKD protocols~\cite{BMM19,VV19,ZM23}, operate according to the principles of quantum mechanics and are physically isolated from one another as well as from any potential adversaries.

\begin{widetext}

\begin{table}[H]
\begin{center}
\scalebox{1}{\begin{tabular}{|c|c|}\hline
\textbf{Symbols} & \textbf{Descriptions}\\\hline
$V$ & subset of vectors of $\mathbb{R}^d$ which satisfies Kochen-Specker theorem\\ \hline
$V^{KS}$ & subset of vectors of $\mathbb{R}^3$ which satisfies Kochen-Specker theorem\\\hline 
$h(\kappa)$ & ternary Shannon entropy with bias $\kappa$\\\hline
$H(P_X)$ & Shannon entropy of the probability distribution $P_X$ \\\hline
$\epsilon_{correct}$ & correct except with probability $\epsilon$\\\hline 
$\epsilon_{secure}$ &  secure except with probability $\epsilon$\\\hline
$\eta$ &  noise tolerance\\\hline
$l$ & length of the active key generated after privacy amplification \\\hline
$\mathcal{N}$& number of input systems\\\hline
$p$ & number of subsystems that are sacrificed for parameter estimation\\\hline
$q$ & size of each block\\\hline
$n$ &number of blocks of size $q$ that are used for the actual computation of the key\\\hline 
$H(\rho_A)$ & Von Neumann entropy of the density operator $\rho_A$\\\hline
$H(A|B)$ & conditional entropy $H(\rho_{AB})-H(\rho_B)$ \\\hline
$H_{min}^{\epsilon}$ & smooth minimum-entropy measures for density operator \\\hline
$H_{max}^{\epsilon}$ & smooth maximum-entropy measures for density operator \\\hline
$r$ & key rate \\\hline
${c}$ & {tunable parameter} \\\hline
\end{tabular}
}\caption{Overview of notations.}
\label{tab:notations}
\end{center}
\end{table}
\end{widetext}

Prior to the protocol's initiation, we assume that Alice and Bob receive $\mathcal{N}$ copies of qutrit entangled states $\frac{1}{\sqrt{3}}(\ket{00}+\ket{11}+\ket{22})$ that are supplied by a third party. The quantum circuit for the preparation of such a state, along with some background on qutrit circuits, is provided in Section \ref{secqtr}.

At this stage, Alice and Bob begin their respective measurement procedures, during which communication between them is strictly prohibited. Upon receiving their inputs, Alice performs a measurement on her share of the entangled state in the basis $B_a = \{\ket{v_1}, \ket{v_2}, \ket{v_a}\} \subset \mathbb{R}^3$, where orthonormalization is applied if necessary. Similarly, Bob measures his share of the entangled state in the basis $B_b = \{\ket{v_\ell}, \ket{v_{b_{1}}}, \ket{v_{b_{2}}}\} \subset \mathbb{R}^3$, also applying orthonormalization if required.

After performing the measurement, Bob publicly announces his output. He declares $0$ if the outcome corresponds to $\ket{v_\ell}$ and $1$ otherwise. According to the quantum winning strategy for the Impossible Colouring game~\cite{BBT05}, Bob obtains $\ket{v_\ell}$ as an outcome if and only if Alice also obtains $\ket{v_\ell}$. Thus, if Bob declares $0$, Alice can infer the chosen vector $\ket{v_\ell}$, and if he declares $1$, the round is discarded.

 After the measurement phase, Alice and Bob discard all instances where Bob did not measure the vector $\ket{v_\ell}$. As a result, approximately $\frac{2\mathcal{N}}{3}$ of the total inputs are discarded. Let the remaining set of rounds be denoted by $\mathcal{A}$. They then randomly select a subset of size $\gamma|\mathcal{A}|$, with $0 < \gamma < 1$, to estimate the success probability of the protocol by checking how often the Impossible Colouring game condition is satisfied. If the observed success probability falls below the threshold $(1 - \eta)$, where $\eta$ represents the acceptable noise level, the protocol is aborted. Otherwise, they proceed to generate a shared secret key using the remaining $(1 - \gamma)|\mathcal{A}|$ rounds.

In the key extraction phase, if Bob selects $\ket{v_\ell} = \ket{v_1}$, both Alice and Bob output $0$, thereby assigning the shared bit value $0$. Similarly, if Bob chooses $\ket{v_\ell} = \ket{v_2}$, both parties output $1$, resulting in the shared bit value $1$. In this way, the raw key is generated from the outcomes of the successful rounds.



 Following the raw key generation, both parties perform classical post-processing, which consists of two main steps: information reconciliation and privacy amplification. During the information reconciliation phase, Alice and Bob exchange information over a public authenticated channel to identify and correct discrepancies in their respective keys. Typically, this involves performing parity checks on short blocks obtained by partitioning the raw key string. As a result, they obtain identical but partially compromised keys. In the subsequent privacy amplification step, they apply a suitable technique, such as a universal hash function or syndrome decoding, to compress the reconciled key into a shorter final key. This final key has significantly reduced information leakage, thereby ensuring that any potential information available to an eavesdropper is rendered negligible. We illustrate our proposed TFDIQKD protocol through an example provided in the next Section \ref{exam}.
\subsection{Example}\label{exam}

Let Alice and Bob want to generate an $ 8$-bit shared secret key. To do this, as explained in the protocol description, Alice and Bob require approximately $24$ entangled states, with $\ket{v_\ell}$ measured on Bob’s side in roughly $24/3=8$ times. Probabilistically, $\ket{v_\ell}$ will not be measured 16 times, so we need to discard approximately 16 rounds; the remaining 8 rounds will provide us desired secret key for both parties. Without loss of generality assume that Bob measures $\ket{v_\ell}=\ket{v_1}$ in rounds $2, 3, 4,$ and $8,$ and $\ket{v_2}$ in the others. By the properties of the Impossible Colouring game described before and in Algorithm \ref{alg3}, their outputs match in each valid round, producing the raw key $10001110$. 

Due to noise, discrepancies may arise. For instance, Bob may obtain $10000110$. To correct this, they perform information reconciliation by dividing the key into blocks (e.g., $4$ bits) and comparing parities. If a mismatch is detected, the block is split to two sub-blocks and so on to locate the error. For example, Alice's last block may have parity $1 \oplus 1 \oplus 1 \oplus 0 = 1$ (odd), while Bob's is $0 \oplus 1 \oplus 1 \oplus 0 = 0$ (even). Therefore, they will split further into 2 length blocks.


After this step, Alice and Bob obtain identical but potentially partially compromised keys. For example, the parities of a block may match even if their bits differ, such as Alice having $1111$ and Bob $0011$. To reduce information leakage, they perform privacy amplification by applying syndrome decoding or a universal hash function, yielding a shorter, secure key. The final key reveals negligible information to any eavesdropper.


\subsection{Algorithm of TFDIQKD protocol}\label{3B}
This section presents the DIQKD protocol based on the Impossible Colouring game. Algorithm \ref{alg:alg1} describes raw key generation and the subsequent classical post-processing to obtain the final key.
 
\begin{algorithm}[H]
 \caption{Proposed TFDIQKD Protocol using Impossible Colouring Game }\label{alg:alg1}
 \textbf{Input:} $\mathcal{N}$ many entangled states $\frac{1}{\sqrt{3}}(\ket{00}+\ket{11}+\ket{22})$ 
 and  noise tolerance $\eta$\\
 \textbf{Output:} Shared common keystream\\
 \textbf{Procedure:}
\begin{enumerate}
    \item State Preparation:
    \begin{itemize}
        \item Alice selects $\ket{v_1},\ket{v_2}$ from $ V^{KS}$ such that they are orthogonal and inputs them into her first two blackboxes. The remaining blackbox of Alice takes $\ket{v_a}$, any vector of $\mathbb{R}^3$ (not necessarily from $ V^{KS}$) as input, which is orthogonal to both $\ket{v_1}$ and $\ket{v_2}$.
         \item Alice sends the first two vectors selected from $ V^{KS}$ to Bob one by one through a secure channel. According to their strategy, the first vector received by Bob is $\ket{v_1}$, and $\ket{v_2}$ will be the next.
         \item Bob chooses randomly one vector received from Alice $\ket{v_\ell}$ as input for the first black box and the other two black boxes take $\ket{v_{b_{1}}},\ket{v_{b_{2}}}$ from $\mathbb{R}^3$,  which is perpendicular to $\ket{v_\ell}.$ 
    \end{itemize}

    \item Measurement: Alice measures her share of the entangled state in basis $B_a=\{\ket{v_1}, \ket{v_2},\ket{v_a}\}$ after orthonormalization and similarly, Bob measures on his share of the entangled state in basis $B_b=\{\ket{v_\ell},\ket{v_{b_{1}}},\ket{v_{b_{2}}}\}$ of $\mathbb{R}^3$ performing necessary normalization. 
    \item Announcement: Bob declares his output publicly $0$ if $\ket{v_\ell}$ measure and $1$ otherwise. Because of the quantum strategy, when Bob declares $0$, then by observing his output, Alice can surely understand what $\ket{v_\ell}$ was chosen by Bob. If Bob declares $1$ as output, then they will discard the step. Let the set $\mathcal{A}$ contain all those cases that are not discarded.
    \item Testing: Alice chooses a random subset $\mathcal{B}\subseteq\mathcal{A}$ of size $\gamma|\mathcal{A}|$ where $0<\gamma\le1$. After measuring the outputs, they discuss their inputs and outputs publicly and estimate the left-hand side of the Bell inequality as mentioned in Equation \ref{ineq}, over $\mathcal{A}\setminus \mathcal{B}$. If this inequality is true, then they abort the protocol. Otherwise, they continue the process for $\mathcal{B}$. 
    \item Key Extraction: In $\mathcal{A}$, when Bob chooses $\ket{v_\ell}$ as $\ket{v_1}$, after measurement Bob gives output $0$ and simultaneously Alice will give output $0$, so their shared common bit becomes $0$, and when Bob chooses  $\ket{v_\ell}$ as $\ket{v_2}$, similarly the bit string will be $1$. Hence, the raw key will be generated.
    \item Classical Post Processing: After the generation of the raw key, Alice and Bob will perform Information Reconciliation and Privacy Amplification to generate the final key with negligible information leakage to ensure that Eve hardly knows anything about it.  
\end{enumerate}
\label{alg1}
 \end{algorithm}
Algorithm \ref{alg1} describes the process of raw key generation and the subsequent classical post-processing to obtain the final key in our proposed DIQKD protocol, which is based on the Impossible Colouring game. Initially, the raw key is generated by Alice and Bob performing quantum measurements based on the pre-shared entangled state and the Impossible Colouring game. The measurement results, denoted as $m_A$ and $m_B$, are then used to construct the raw key. A Bell test is applied to verify the violation of Bell's inequality, ensuring the device-independent nature of the protocol. After discarding inconsistent results that do not satisfy the Bell test, classical post-processing steps, such as error correction and privacy amplification, are applied to the raw key. The final outcome is a secure, shared secret key between Alice and Bob.

 In Figure \ref{Fi}, we depicted the pseudocode for the generic QKD protocol, where the most crucial step was Raw Key generation. The raw key generation was done using a qutrit entangled state and the pseudocode for the same is depicted in Algorithm \ref{alg3}. 
 
\begin{algorithm}[H]
    \caption{Qutrit Entanglement-based Raw Key Distribution}
    \begin{algorithmic}
        \Require\\
        $\bullet$ $\mathcal{N}$-many entangled state $\ket{\psi} = \frac{1}{\sqrt{3}}(\ket{00}+\ket{11}+\ket{22})$ [Bases for encoding]\\$\bullet$ Noise tolerance = $\eta$
        
        \Ensure Raw key without involvement of any devices
        
        \Function {Entanglement Based RawKeyDistribution}{}
        \State $i$=1
            \While{$i\le \mathcal{N}$}\\
                Alice chooses $B_a=\{\ket{v_1},\ket{v_2},\ket{v_a}\}$\\
                Bob chooses $B_b=\{\ket{v_\ell},\ket{v_{b_1}},\ket{v_{b_2}}\}$, with $v_l\xleftarrow{\$}\{v_1,v_2\}$\\
                Alice and Bob measure using their basis\\
                \If{$\ket{v_\ell}$ not measured}
                \State \textbf{return} $(\perp,\perp)$ and \textbf{discard}
                \EndIf
                    \If{$\ket{v_\ell}$ is $\ket{v_1}$}
                    \State \textbf{return} 0
                    \EndIf
                    \If{$\ket{v_\ell}$ is $\ket{v_2}$}
                    \State \textbf{return} 1
                    \EndIf
            \EndWhile{$i:=i+1$}     
       \State \Return \textbf{(K$_A^R$,K$_B^R$)}          \EndFunction
    \end{algorithmic}
    \label{alg3}
\end{algorithm}

We shall now provide the quantum circuit which is required to construct the entangled state $\frac{\ket{00}+\ket{11}+\ket{22}}{\sqrt{3}}$ and other similar states also known as $3$-dimensional Bell basis states \cite{SL09}.

\subsection{Quantum circuits for preparation of ternary Bell basis}\label{secqtr}
In this section, we design qutrit quantum circuits that prepare the maximally entangled states like $
\ket{\varphi_0} = \frac{\ket{00} + \ket{11} + \ket{22}}{\sqrt{3}}$, which serves as a fundamental resource for our protocol. This state is an element of the $3$-dimensional Bell basis, a natural generalization of the standard Bell basis to qutrit systems \cite{SL09}. The complete $3$-dimensional Bell basis is defined as follows:
\begin{eqnarray*}
    \ket{\varphi_0} &= \frac{\ket{00}+\ket{11}+\ket{22}}{\sqrt{3}}, \\
    \ket{\varphi_1} &= \frac{\ket{00}+\omega\ket{11}+\omega^2\ket{22}}{\sqrt{3}},\\
    \ket{\varphi_2} &= \frac{\ket{00}+\omega^2\ket{11}+\omega\ket{22}}{\sqrt{3}}, \\
    \ket{\varphi_3} &= \frac{\ket{01}+\ket{12}+\ket{20}}{\sqrt{3}}, \\
    \ket{\varphi_4} &= \frac{\ket{01}+\omega\ket{12}+\omega^2\ket{20}}{\sqrt{3}}, \\
    \ket{\varphi_5} &= \frac{\ket{01}+\omega^2\ket{12}+\omega\ket{20}}{\sqrt{3}}, \\
    \ket{\varphi_6} &= \frac{\ket{02}+\ket{10}+\ket{21}}{\sqrt{3}}, \\
    \ket{\varphi_7} &= \frac{\ket{02}+\omega\ket{10}+\omega^2\ket{21}}{\sqrt{3}}, \\
    \ket{\varphi_8} &= \frac{\ket{02}+\omega^2\ket{10}+\omega\ket{21}}{\sqrt{3}}.
\end{eqnarray*}
Here, $\omega = e^{2\pi i/3}$ is the principal cube root of unity. These states form an orthonormal basis for the joint qutrit Hilbert space $\mathbb{C}^3 \otimes \mathbb{C}^3$ and are frequently employed in high-dimensional entanglement-based quantum communication protocols. This set of maximally entangled two-qutrit states serves as the fundamental resource for implementing the quantum correlations required by our protocol. In order to construct a qutrit circuit for the preparation of the state, we shall discuss a bit about qutrit gates. From \cite{sarkar2020periodicity}, we consider the following set of single qutrit gates \begin{eqnarray}\label{xgates}
&& X_{0,1}=\bmatrix{0&1&0\\1&0&0\\0&0&1}, X_{1,2}=\bmatrix{1&0&0\\0&0&1\\0&1&0},\\\nonumber && X_{0,2}=\bmatrix{0&0&1\\0&1&0\\1&0&0}, X_{+1}=\bmatrix{0&0&1\\1&0&0\\0&1&0},\\ &&X_{+2}=\bmatrix{0&1&0\\0&0&1\\1&0&0}.\nonumber
\end{eqnarray} 
Obviously, the $X_{p,q}$ gate maps $\ket{p}_3$ (respectively, $\ket{q}_3$) to $\ket{q}_3$ (respectively, $\ket{p}_3$), where $p, q \in \{0,1,2\}$. The $X_{+a}$ gate, on the other hand, defines a linear transformation from $\ket{p}_3$ to $\ket{(a + p) \bmod 3}_3$. We refer to these single-qutrit gates collectively as \textit{qutrit-$X$ gates}, where $X \in \{X_{0,1}, X_{1,2}, X_{0,2}, X_{+1}, X_{+2}\}$. We also consider the single qutrit Hadamard gate \cite{sarkar2020periodicity} $H_3=\bmatrix{1&1&1\\1&\omega&\omega^2\\1&\omega^2&\omega}$ where $\omega=\exp{\frac{2\pi i}{3}}$ i.e. a complex root of unity.

Also, following \cite{sarkar2020periodicity}, we have the {two-qutrit controlled-$X$ gates} represented by a quantum circuit given by
\begin{eqnarray}
{\Qcircuit @C=1em @R=.7em {
 &\lstick{A}&\qw& \gate{\raisebox{-1pt}{\textcircled{\raisebox{1pt} {$\alpha$}}} }& \qw\\
 &\lstick{B}&\qw& \gate{X} \qwx[-1]& \qw\\}}    
\end{eqnarray}
Obviously, this gate applies a one-qutrit $X$ gate to the target (second) qutrit when the control (first) qutrit is in the state $\alpha \in \{0,1,2\}$. In particular, when $\alpha = 2$, such a gate is known as the Muthukrishnan–Stroud gate \cite{MS00}. For further details on qutrit gates, see \cite{sarkar2020periodicity}. With this background, we shall construct a quantum circuit involving qutrits in order to prepare our $3$-dimensional Bell states. Let us consider the following circuit.

\begin{eqnarray}
{\Qcircuit @C=1em @R=.7em {
 &\lstick{q_0}&\qw&\gate{H_3}& \gate{\raisebox{-1pt}{\textcircled{\raisebox{-.75pt} {1}}} }& \qw& \gate{\raisebox{-1pt}{\textcircled{\raisebox{-.75pt} {2}}} }&\qw\\
 &\lstick{q_1}&\qw&\gate{U}& \gate{X_{+1}} \qwx[-1]& \qw& \gate{X_{+2}} \qwx[-1]&\qw\\}}    
\end{eqnarray}

It is straightforward to verify that if $U = I$ (the identity operator), then starting from the initial two-qutrit state $\ket{00}$ (i.e., $q_0 = q_1 = 0$), the output of the circuit is the maximally entangled state $\ket{\varphi_0}$. Similarly, if $U = X_{+1}$, the input $\ket{00}$ results in the output state $\ket{\varphi_3}$, and if $U = X_{+2}$, the output becomes $\ket{\varphi_6}$.

For the input $\ket{10}$, applying $U = I$ yields $\ket{\varphi_1}$, applying $U = X_{+1}$ gives $\ket{\varphi_4}$, and applying $U = X_{+2}$ produces $\ket{\varphi_7}$.

Finally, for the input $\ket{20}$, the circuit outputs $\ket{\varphi_2}$ when $U = I$, $\ket{\varphi_5}$ when $U = X_{+1}$, and $\ket{\varphi_8}$ when $U = X_{+2}$.

In the following section, we shall discuss the security analysis of our QKD protocol.



\section{Security Analysis of Our Proposed Protocol}\label{sec4}

In this section, we shall exhibit the security of our QKD protocol. Although the characterization of the quantum state and the measurement devices is not required for the security of DIQKD, certain assumptions and security notions are still necessary to establish its security rigorously. The standard assumptions underlying the security analysis of our proposed protocols are outlined below.
\begin{enumerate}

    \item[i.] \textbf{Memoryless Device:} The devices must operate in accordance with the principles of quantum mechanics, with each use being independent of previous uses, and must exhibit consistent behaviour across all trials.
    \item[ii.] \textbf{Secured Laboratory:} Alice’s and Bob’s laboratories are assumed to be completely secure, preventing any leakage of confidential information. Furthermore, the laboratories are physically separated. This enables the modeling of their systems as jointly distributed quantum systems.
    \item[iii.] \textbf{Random Seeds:} We assume that both parties have access to uniformly random seeds, which are typically generated locally using Quantum Random Number Generators (QRNGs) \cite{MYCQZ16} within their respective laboratories.
    \item[iv.] \textbf{Authenticated Communication Channel:} It is assumed that Alice and Bob communicate over a public, authenticated channel. All information transmitted through this channel is considered publicly accessible and treated as part of the protocol’s output. By appropriately labeling each classical message, information-theoretic techniques can be employed to ensure authentication of the classical communication channel. 
    \end{enumerate}

In general, a QKD protocol consists of two main phases. Firstly, both parties share a maximally entangled state through a secure quantum channel. Next, they apply a key distillation scheme to extract the final secret key. Consequently, for the security analysis of QKD, it suffices to establish the security of the underlying key distillation process. A schematic overview of the complete key distillation scheme is provided in Table \ref{Figure1}.

Before proceeding with the security analysis, we must ensure that the key generation scheme is correct and device-independent. Specifically, if an adversary attempts to construct a correlated system and interact with the protocol in parallel with the legitimate parties, their probability of winning the game should remain negligibly small. Further, the generated raw key should be distinguishable, barring a small fraction attributable to noise tolerance.

Therefore, prior to initiating the final key security analysis using the smooth minimum-entropy framework proposed by Renner \cite{RR08, PR22}, we demonstrate in Section \ref{subsec:3.1} that our scheme is $\epsilon$-correct. Section \ref{subsec:3.2} establishes the device independence of our protocol, and Section \ref{subsec:3.3} ensures the security of the raw key.


\subsection{Correctness of Our Protocol}
\label{subsec:3.1}
In this section, we show the correctness of our protocol using the definition of $\epsilon$-correct. 
\begin{definition}
\label{def1}
(\textbf{$\epsilon$-correct})
A quantum key distribution protocol is $ \epsilon$-correct ~\cite{PR22,BMM19} if given a security parameter $\epsilon \ge 0$, Alice and Bob agree on a $k$-bit key $K \in \{0, 1\}^k$, except with some failure probability at most $\epsilon$ i.e. both parties achieve the key streams $K_A$, $K_B$ respectively, such that
Pr($K_A \ne K_B$) $\le \epsilon$.
\end{definition}

 We now show the correctness of our proposed DIQKD protocol by establishing a suitable $\epsilon$.
\begin{theorem}\label{thmiv2}
    The proposed QKD scheme guarantees an $\epsilon$-correctness level, which is defined as $\epsilon_{\text{correct}}=[1-(1-\eta)^k]$.
Here, `$k$'  represents the length of the entire raw key, and `$\eta$' is the tolerable noise parameter.
\end{theorem}
\pf See Appendix \ref{appendix:B}. $\hfill\square$



\subsection{Fully Device Independency of Our Protocol}
\label{subsec:3.2}
We now turn our attention to the security of the overall protocol, which can be analyzed in two parts: the security of the fully device-independent testing phase and the security of the quantum key distillation phase.

In generic models of FDIQKD protocols, Bell nonlocality serves as a fundamental principle for verifying device independence \cite{bell14,PR22}. The key idea involves the use of Bell inequalities \cite{bell14,GNM07}, which are constraints that must be satisfied by any theories based on local hidden variables. A violation of these inequalities indicates the presence of correlations that cannot be explained by classical means, thereby revealing inherently quantum and nonlocal behavior. Importantly, such violations can be observed solely through input-output statistics, without any assumptions about the internal workings of the devices involved. This feature allows Bell inequality violations to serve as a powerful tool for establishing device independence, guaranteeing that the security or performance of a protocol is independent of the specific physical implementation.


Our protocol is based on the impossible colouring game, a pseudo-telepathy game constructed using the Kochen-Specker theorem. This theorem explores the impossibility of assigning deterministic outcomes, interpreted as a $\{0,1\}$ colouring, to a certain set of vectors in $\mathbb{R}^d$, under the following two constraints: (i) only one vector in any orthogonal pair may be assigned the value 1, and (ii) at least one vector in every orthogonal triplet must be assigned the value 1. Such sets are known as Kochen-Specker sets. The impossibility of such a colouring illustrates a fundamental feature of quantum mechanics known as contextuality \cite{BC22}.

Contextuality refers to the phenomenon in which the outcome of measuring an observable depends not solely on the system, but also on the specific set of compatible measurements performed alongside it, referred to as the context. This feature is fundamental to the impossible colouring game. In this game, two players receive orthogonal contexts drawn from a Kochen-Specker set and must assign values of 0 or 1 to each vector. They are promised that one vector in their input is identical to a vector in the other player’s input, although they do not know which one. Their task is to assign the value 1 to this shared vector. There is a nonzero probability that both players will select the same vector. However, any classical deterministic strategy must assign definite $\{0,1\}$ values to the vectors, thereby violating the Kochen-Specker constraints and contradicting non-contextuality.

In contrast, a quantum strategy allows the players to always succeed. Their responses abides by the Kochen-Specker constraints and yield the shared vector, despite the impossibility of doing so in any classical non-contextual hidden variable model. This ability to maintain global consistency through local, context-dependent outcomes is not a technical artifact. It provides a concrete manifestation of contextuality as a physical principle.
 The fundamental challenge with contextuality, despite its close connection to nonclassical behavior, lies in the absence of a well-defined Bell inequality. Although it was commonly believed that contextuality implies Bell nonlocality, there existed no formal mathematical limit or inequality capable of capturing this implication. In particular, there was no Bell-type inequality whose violation could demonstrate that contextual behavior leads to nonlocal correlations.

Several works have attempted to bridge this conceptual gap. For instance, Kunkri et al. \cite{KKGR07}, Renner et al. \cite{RW04}, and Zimba et al. \cite{ZP93} explored the relationship by simulating Kochen-Specker-type contextual scenarios using nonlocal boxes or alternative geometric constructions. These approaches established structural analogies between contextuality and nonlocality but did not produce a concrete, experimentally testable inequality.

This gap was addressed in a series of papers by Cabello et al.\cite{cabello15,cabello21,cabello2021bell}, who showed that any Kochen-Specker contextuality scenario can be mapped to a bipartite nonlocality scenario. Significantly, the authors provided a constructive method for this transformation, including a procedure to derive a Bell inequality from the original contextual setup. This framework allows contextuality-based games, such as the impossible colouring game, to be treated within the standard Bell framework, enabling formal, device-independent proofs based on violations of well-defined nonlocal constraints.

\begin{figure}[H]
  \centering
  \includegraphics[width=9.0cm]{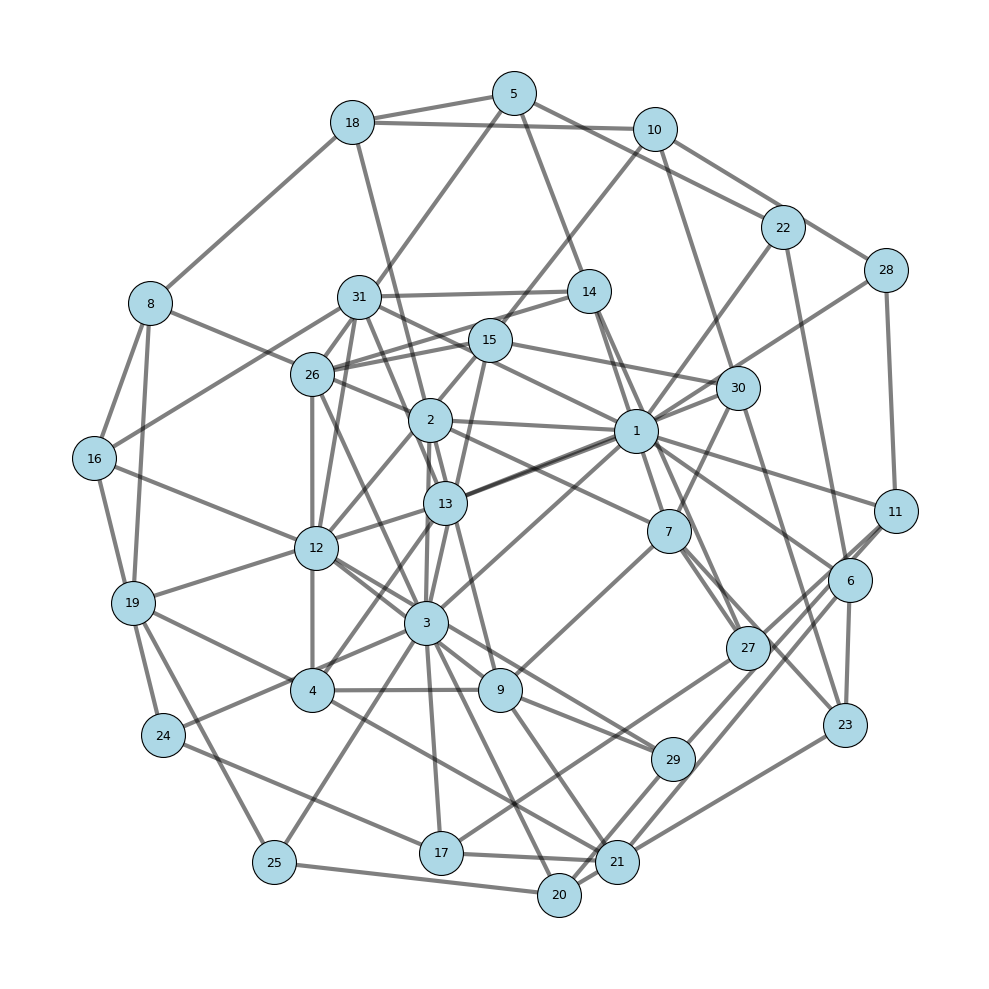}\\
    \caption{The orthogonality graph of the Conway-Kochen Set of 31 vectors, denoted as $\mathcal{G}$. Note that an edge exists between two nodes if and only if the vectors corresponding to the nodes are orthogonal to each other. Extensive details about the vectors are provided in Table \ref{graph_table}.}
    \label{sicgraph}
\end{figure}

The algorithm as stated by Cabello et al. \cite{cabello21} used to perform this mapping is detailed, as follows.
\begin{enumerate}
    \item First, check whether the Kochen--Specker set $S = \{\ket{v_1}, \ket{v_2}, \ldots, \ket{v_n}\}$ is a state-independent contextuality (SI-C) set. To do this, construct the {orthogonality graph} $G$, where each vector in $S$ corresponds to a node, and an edge is drawn between two nodes if the associated vectors are orthogonal. Then, verify whether both the chromatic number $\chi(G)$ and the fractional chromatic number $\chi_f(G)$ are strictly greater than the dimension $d$ of the vectors. This condition is necessary for the set to exhibit state-independent contextuality.
    \begin{enumerate}

        \item If the check fails, it implies the set is state-dependent contextual (SD-C). If it is an SD-C set, it can be converted to a SI-C set as stated in this article \cite{cabello21}.
        \item If the check is true, this still does not confirm the set is SI-C. To confirm this, there must exist non-negative numbers $w'_1, w'_2, \ldots, w'_n$ and a number $y'$ such that $0 \leq y' < 1$, satisfying the following conditions: for all independent sets $\mathcal{I}$ in the graph $G$, we have $\displaystyle\sum_{i \in \mathcal{I}} w_i \leq y$, and $\displaystyle\left(\sum_i w_i \Pi_i\right) - \mathbb{I}$ is positive semi-definite where $\mathbb{I}$ is the $d\times d$ Identity operator. If these conditions hold, the set is indeed SI-C. Otherwise, repeat the process described earlier to obtain an SI-C set. Maximize $\sum_i w'_i$ to obtain a unique set of weights $\{w_i\}_{i=1}^n$ and the corresponding number $y$.
    
    \end{enumerate}
    
    Once these steps are done, we obtain a new graph $\mathcal{G}$ generated from $G$ and a new set $S'$ derived from $S$.
    In our case, based on the Conway–Kochen vector set \cite{peres97}, we obtain an orthogonality graph $\mathcal{G}$ as shown in Figure~\ref{sicgraph}. We find that $\chi(\mathcal{G}) = 4$ and the fractional chromatic number $\chi_f(\mathcal{G}) = 3.2$, confirming that the set is not state-dependent contextual (SD-C). To verify that it is state-independent contextual (SI-C), we check for the existence of a tuple $(w_1, \ldots, w_{31}, y)$ satisfying the conditions of step (1b). Once a feasible tuple is found, we maximize $\sum w_i$ to obtain a unique solution.
    
    Our implementation uses Python, specifically the libraries \texttt{numpy}, \texttt{networkx}, and \texttt{cvxpy}, executed in a Google Colab environment with a 2.4 GHz CPU and 24 GB RAM. Since the verification involves both semidefinite and linear programming, we use the \texttt{Splitting Cone Solver (SCS)} package. The solver returns a valid solution $(w_i)_{i=1}^{31}$ with $y = 0.999993642346489$.
    \item Set the rank-one projections as $\Pi_i := \ket{v_i}\bra{v_i}$. 
From the previous step, besides obtaining $\mathcal{G}$ and $S'$, we also obtain another set $\{w_1, w_2, \ldots, w_n, y\}$. 
We assign the weights associated with each $\Pi_i$ and its corresponding node in $\mathcal{G}$ as $W_i := \dfrac{w_i}{y}$. 
Note that $\sum_i W_i > d$. 
We define $\alpha(\mathcal{G}, W)$ to be the weighted independence number of the graph $\mathcal{G}$, where the $i$-th node is assigned the weight $W_i$, and $W := \{W_i\}_{i=1}^n$.

    
    Now, denoting $\vartheta(G,W)$ as the weighted Lov\'{a}sz number \cite{cabello14}, we get the following contextuality inequality:
    \small\begin{eqnarray*}
        &&\sum_{i\in V(\mathcal{G})} W_iP(\Pi_i=1) - \sum_{(i,j)\in E(\mathcal{G})} \max(W_i,W_j)P(\Pi_i=1,\Pi_j=1)\\&&\overset{NCHV}{\le}\alpha(\mathcal{G},\{W_i\}_{i=1}^d)\overset{QT}\leq\vartheta(G,\{W_i\}_{i=1}^n).
    \end{eqnarray*}\normalsize
Note that ``NCHV" in this inequality stands for Non-Contextual Hidden Variable Theories.
    We calculate normalized weights $W_i := \dfrac{w_i}{0.999993642346489}$. The $31$ rays $v_i$ from the Conway–Kochen set, alongside their corresponding normalized weights $W_i$ is mentioned in Table \ref{graph_table}. We compute the weighted independence number $\alpha(\mathcal{G}, \{W_i\}_{i=1}^{31})$. It is of note that the independent set $\mathcal{I}' = \{2, 4, 6, 11, 12, 14, 17, 20, 30\}$ yields the maximum total weight. Therefore, the weighted independence number is given by $\alpha(\mathcal{G}, \{W_i\}) = \sum_{i \in \mathcal{I}'} W_i = 1.0000112242568169$. We skip the calculation for the weighted Lov\'{a}sz number, as it is computationally cumbersome, and it is enough to focus on the classical bound for the verification process. 

    \item We prepare our entangled two-state qudit system $\ket{\psi}:=\displaystyle\sum_{k=0}^{d-1}\ket{k}\ket{k}$ and set Alice's and Bob's vector set as $S'$ and $\bar{S'}$, where $\bar{S'}$ is the complex conjugate of the elements of $S'$. Now we have the following Bell-inequality:
    \small\begin{eqnarray*}
    &&\sum_{i\in V(\mathcal{G})}W_iP\left(\Pi_i^A=1,\Pi_i^B=1\right)-\Bigg(\sum_{(i,j)\in E(\mathcal{G})}\frac{\max(W_i,W_j)}2 \\&&\quad \left[P\left(\Pi_i^A=1,\Pi_j^B=1\right)+P\left(\Pi_j^A=1,\Pi_i^B=1\right)\right]\bigg)\\&&\overset{LHV}{\leq} \alpha(\mathcal{G},\{W_i\}_{i=1}^n).
\end{eqnarray*}\normalsize
Note that $P (\Pi^A_i = 1, \Pi^B_j = 1)$ is the probability that Alice obtains outcome 1 for measurement $\Pi_i$ on her particle and Bob obtains outcome 1 for measurement $\Pi_j$ on his particle, and that ``LHV" stands for Local Hidden Variable Theories.
\end{enumerate}

Hence, from the discussions so far, we get the following inequality.
\begin{eqnarray}\label{ineq}
    &&\sum_{i\in V(\mathcal{G})}w_iP\left(\Pi_i^A=1,\Pi_i^B=1\right)-(\sum_{(i,j)\in E(\mathcal{G})}\frac{\max(w_i,w_j)}2\nonumber\\&&  \cdot\left[P\left(\Pi_i^A=1,\Pi_j^B=1)+P\left(\Pi_j^A=1,\Pi_i^B=1\right)\right]\nonumber\right)\\&&\overset{LHV}{\leq} 1.0000112242568169 \, .
\end{eqnarray}\normalsize

Here, the calculations have been carried out with respect to the orthogonality graph of the Conway-Kochen Set of $31$ vectors, which is depicted in Figure \ref{sicgraph}. 

\begin{table}[htb]
    \centering
    \begin{tabular}{c c c}
\hline
Node $i$ & Rays $v_i$ & Corresponding weights $W_i$ \\
\hline
$1$ & $(1 , 0 , 0)^T$ & $0.25799809370143945$ \\
$2$ & $(0 , 0 , 1)^T$ & $0.19999263759649782$ \\
$3$ & $(0 , 1 , 0)^T$ & $0.2579980937016148$ \\
$4$ & $(-1 , 1 , -1)^T$ & $0.12473071346150894$ \\
$5$ & $(-1 , 2 , -1)^T$ & $0.04198675640899053$ \\
$6$ & $(0 , 2 , -1)^T$ & $0.07527393320356464$ \\
$7$ & $(-1 , 1 , 0)^T$ & $0.19999975889766655$ \\
$8$ & $(-1 , 2 , 0)^T$ & $0.04198868808244714$ \\
$9$ & $(1 , 1 , 0)^T$ & $0.1999997588976661$ \\
$10$ & $(-1 , 2 , 1)^T$ & $0.0419867564089987$ \\
$11$ & $(0 , 2 , 1)^T$ & $0.0752739332035787$ \\
$12$ & $(-1 , 1 , 1)^T$ & $0.12473071346150441$ \\
$13$ & $(0 , 1 , 1)^T$ & $0.14075323938552295$ \\
$14$ & $(1 , 1 , 1)^T$ & $0.12473071346150931$ \\
$15$ & $(1 , 0 , 1)^T$ & $0.1407532393865908$ \\
$16$ & $(2 , 1 , 1)^T$ & $0.04198675640808494$ \\
$17$ & $(2 , 0 , 1)^T$ & $0.07527393320356927$ \\
$18$ & $(2 , 1 , 0)^T$ & $0.041988688083343585$ \\
$19$ & $(2 , 1 , -1)^T$ & $0.04198675640808829$ \\
$20$ & $(2 , 0 , -1)^T$ & $0.07527393320357453$ \\
$21$ & $(-1 , 1 , 2)^T$ & $0.07527928769251237$ \\
$22$ & $(0 , 1 , 2)^T$ & $0.04198226094368069$ \\
$23$ & $(1 , 1 , 2)^T$ & $0.07527928769251353$ \\
$24$ & $(-1 , 0 , 2)^T$ & $0.0419822609427853$ \\
$25$ & $(1 , 0 , 2)^T$ & $0.041982260942795356$ \\
$26$ & $(-1 , 0 , 1)^T$ & $0.1407532393865908$ \\
$27$ & $(-1 , -1 , 2)^T$ & $0.07527928769251417$ \\
$28$ & $(0 , -1 , 2)^T$ & $0.04198226094369912$ \\
$29$ & $(1 , -1 , 2)^T$ & $0.07527928769251518$ \\
$30$ & $(-1 , -1 , 1)^T$ & $0.12473071346150935$ \\
$31$ & $(0 , -1 , 1)^T$ & $0.14075323938552078$ \\
\hline
\end{tabular}

    \caption{The table of nodes and their corresponding rays and weights of graph $\mathcal{G}$ as depicted in Figure \ref{sicgraph}}
    \label{graph_table}.
\end{table}

Further, in order to ensure device independence in the Impossible Colouring game, it is necessary to understand the observed success probability, particularly to distinguish classical strategies from quantum ones. Since classical strategies are bounded by a dimension-dependent success threshold, any observed violation of this bound certifies nonclassical behavior purely from statistics, without relying on assumptions about the devices.

\begin{theorem}\label{Thd}
Let a $d$-dimensional Impossible Colouring game be played between two parties. Then the maximum success probability achievable by any classical strategy is $\dfrac{1}{d}$.
\end{theorem}

\noindent\textit{Proof.} See Appendix~\ref{appendix:C}. \hfill$\square$

For our protocol, we now state the specific instance of Theorem~\ref{Thd} for the case $d = 3$.

\begin{corollary}\label{cor44}
     When a $3$-dimensional Impossible Colouring game is played among $2$ parties, then the best possible approach of winning this game for these $2$ parties will have a probability of $\dfrac{1}{3}$.
\end{corollary}

It can be seen that for the quantum case, from the proof of Corollary \ref{cor44} (\ref{appendix:B}), that the probability $$P(\Pi_i^A=1,\Pi_j^B=1)=\begin{cases}
    \dfrac13, & \text{ if }i=j\\
    0, & \text{ otherwise}
\end{cases}$$

Thus, the left-hand side of the inequality  simplifies to $\dfrac{1}{3} \sum_i W_i = 1.066663494447466$ , which is clearly greater than the right-hand side of the inequality mentioned in Equation \ref{ineq}. Therefore, the quantum case violates this inequality, thereby establishing the device independence of the proposed protocol.

In a generic classical noncontextual or local hidden-variable model with their associated weighted graph \cite{cabello14}, each event is assigned a predetermined value: either it occurs, which is counted as 1, or it does not (to be counted as 0). Since mutually exclusive events cannot both occur, such models correspond to choosing independent sets in the exclusivity graph \( G \). The maximum total weight attainable in any classical strategy is therefore upper bounded by the \emph{weighted independence number} \( \alpha(G, w) \).

Further, as shown in some works in the literature such as \cite{BMM19}, if Alice and Bob each receive three particles from an $\mathcal{N}$-partite quantum state, with the remaining $\mathcal{N} - 3$ held by an adversary (Eve), the protocol can still exhibit device independence provided the observed success probability $p$ in the Impossible Colouring game lies in $[1 - \eta, 1]$. Since this game is a pseudo-telepathy game, such performance cannot be achieved by classical or adversarially correlated strategies. As we will later see, the threshold $\eta$ is $18.11\%$, and the result in Corollary~\ref{cor44} implies that any substantial correlation with Eve would be detectable during the testing phase. This provides an alternative perspective, validating the conclusion that our protocol achieves device-independent quantum key distribution.

\subsection{Security of Raw Key}
\label{subsec:3.3}
 In this section, we focus on the key aspect of the security analysis, namely the security of the key distillation phase. In a quantum key distillation protocol, Alice and Bob receive inputs from $\mathcal{N}$-dimensional Hilbert spaces $\mathcal{H}_A^{\otimes \mathcal{N}}$ and $\mathcal{H}_B^{\otimes \mathcal{N}}$, respectively, and proceed through the following subprotocols:
\begin{itemize}
    \item[i.] Random permutation of the subsystems,
    \item[ii.] Parameter estimation,
    \item[iii.] Block-wise measurement, and
    \item[iv.] Classical post-processing.
\end{itemize}
A schematic representation of the key distillation procedure, detailing each of the above subprotocols, is {provided in Table \ref{Figure1}}.


In classical cryptographic protocols, security is usually defined in terms of random systems by measuring their distance from an ideal system, which is secure by definition \cite{BMM19}. The ideal system for a classical Key Distribution scheme generates a uniform random key at one end, with all interfaces being completely independent of one another. Since the adversary must guess uniformly from the set of all possible keys, the key is considered secure by construction, since each possible string occurs with equal probability. A key is therefore regarded as secure if no significant distinction can be made between the real and ideal systems that generate it.

In the quantum setting, however, the adversary (Eve) may prepare a correlated quantum system to gain information about the measurements and corresponding raw key bits shared between the parties. If Eve's system is uncorrelated with the shared quantum state of the parties, then she cannot obtain any information. In this scenario, the overall system is considered secure if Eve’s subsystem is uncorrelated with that of Alice and Bob, and the joint state appears to be maximally mixed from Eve’s perspective. According to the key distillation phase, the key bits generated after the block-wise measurement phase are referred to as the raw key.


In general, the closeness measure is defined using the definition of statistical distance as 
\begin{definition}
\label{def2}
(\textbf{Statistical Distance:}) \cite{BMM19}
Let \texttt{X}$_0$ and \texttt{X}$_1$ be two
random variables over a finite set $\mathcal{C}$. Then the statistical distance $\Delta(\texttt{X}_0, \texttt{X}_1)$ is defined as:
$$\Delta(\texttt{X}_0, \texttt{X}_1)= \frac{1}{2}\sum\limits_{x \in \mathcal{C}}|[\Pr(\texttt{X}_0=x)-\Pr(\texttt{X}_1=x)]|.$$
\end{definition}
This notion of quantifying closeness in the classical setting using statistical distance leads to the following theorem, which illustrates the security of the generated raw key. Further, we shall also provide the formal definition of distinguishability in the context of this paper. We note that the statistical distance draws parallel with Total Variation Distance (TVD) of probability measures \cite{LP17}.

 \begin{definition}\label{eta}
    The raw key bit generated in any protocol is said to be $\eta$-distinguishable if for any Probabilistic Polynomial Time (PPT) adversary, $|\Pr\mbox{(adversary guesses raw bit correctly)}-\Pr\mbox{(adversary guesses random bit correctly)}| \leq \eta$ 
 \end{definition}

Then we have the following theorem. 
\begin{theorem}\label{iv7}
In the proposed protocol, each raw key bit that is generated is at most $\eta$-distinguishable from a randomly generated bit, where $\eta$ represents the tolerable noise parameter for the protocol.
\end{theorem}

\pf  See Appendix \ref{appendix:D}. $\hfill\square$
\subsection{Security of Active (Final) Key}
\label{IV.d}

In this section, we shall provide the length of the active key and subsequently derive the key rate. The Lo-Chau security proof \cite{LCF99} was one of the first to rigorously establish the theoretical security of QKDs, specifically for the BB84 protocol \cite{BB84}, and was later extended to DIQKD protocols. It introduced the reduction of QKD to entanglement purification protocols. Following this, Shor and Preskill \cite{SPJ00} simplified the Lo-Chau proof by connecting it to Calderbank-Shor-Steane (CSS) codes \cite{CS96,S96}. These proofs were groundbreaking and laid the foundation for QKD security. Renner’s proof \cite{RR08} offers another comprehensive, realistic, and practically applicable security framework. It addresses a wider range of practical issues, provides finite-key security guarantees, and ensures composable security. In 2020, Tsurumaru \cite{tsurumaru20} simplified the idea of QKD security by showing a mathematical equivalence between the security proofs following Mayers(?)-Shor-Preskill and that of Renner's using the Leftover Hashing Lemma. In this paper, we will follow Renner's approach in validating security and estimating the key rate of our given scheme.  It is to be noted that our proposed scheme is inherently \textit{one-way} \cite{RR08} and does not require noisy preprocessing. Specifically, after the measurement of the entangled subsystems, Alice and Bob proceed directly to an information reconciliation protocol, wherein Bob computes an estimate of Alice’s raw key using the information exchanged over an authenticated classical channel. This streamlined design avoids additional classical interaction and preprocessing steps, thereby reducing operational complexity. As a result, the protocol remains efficient and practical, particularly in low-latency or high-noise quantum network environments.

\begin{widetext}

\begin{center}

\begin{table}[H]
\begin{tabular}{l}\hline
{Device Independent Quantum Key Distillation Scheme for QKD$_{\texttt{PE,BM,CPP}}$}\\
\hline
Parameters:\\
\texttt{PE:} \hspace{1cm} Parameter estimation\\
\texttt{BM:} \hspace{1cm} Block wise Measurement\\
\texttt{CPP:} \hspace{0.8cm} Classical Post-Processing\\
$\mathcal{N}$: \hspace{1.1cm} Number of input systems ($\mathcal{N}\ge p+qn$) (The symbols $p, q,\text{and } n$ follow from Table \ref{tab:notations})\\
   \begin{tabular}{ccc}
\textbf{Alice}&&\textbf{Bob}\\

\vspace{0.5cm}

   Input system: $\mathcal{H}_A^{\otimes \mathcal{N}}$&&Input system: $\mathcal{H}_B^{\otimes \mathcal{N}}$\\ 
   
   \vspace{0.5cm}
   
  Random permutation of subsystem &$ \xleftrightarrow{\pi}$& Random permutation of subsystem \\
  \vspace{0.5 cm}
  
  $\mathcal{H}_A^{\otimes p}$&$\xleftrightarrow{\text{\texttt{PE}}}$& $\mathcal{H}_B^{\otimes p} \longrightarrow \text{(accept/abort)}$\\
  
  \vspace{0.5cm}
  
 $(\mathcal{H}_A^{\otimes q})^{\otimes n}$&$ \xleftrightarrow{\text{\texttt{BM}}^{\otimes n}}$& $(\mathcal{H}_B^{\otimes q})^{\otimes n}$\\
\vspace{1cm}

$(u_1,u_2,\cdots, u_n)\rightarrow K_{A}^{R}$ &$\xleftrightarrow{\text{\texttt{CPP}}}$& $(v_1,v_2,\cdots, v_n)\rightarrow K_{B}^{R}$\\

\vspace{0.5cm}

 
$\text{output } K_{A}^{F}$&&$ \text{output } K_{B}^{F}$

   \end{tabular}

\end{tabular}
   \caption{Schematic overview of the proposed device-independent key distillation protocol.}\label{Figure1}
\end{table}   
\end{center}
\end{widetext}

We shall now determine the key rate $r$ based on the work of Portman and Renner \cite{RR08,PR22}. We make use of the tolerable noise parameter $\eta$ which is equivalent to quantum bit error rate in \cite{RR08,PR22}. In this paper, $\eta$ represents the Quantum Trit Error Rate (QTER). In this context, we focus on the modular approach to security analysis proposed by Portman and Renner~\cite{PR22} in 2022. All security proofs share a common goal, which is to establish a relationship between the information available to authorized parties and the information that could have been obtained by Eve. In this regard, analysis of the key rate of the protocol is fundamental to its computation. The key rate of the quantum key distribution scheme is provided as follows.
\begin{eqnarray*}
    \text{Key rate} = \min_{\sigma_{AB} \in \Gamma} H(X|E) - H(X|Y),
\end{eqnarray*}
where the set $\Gamma$ is defined as
\begin{eqnarray*}
    \Gamma = \left\{ \sigma_{AB} : P_W^{\sigma_{AB}} \in \mathcal{Q} \right\}.
\end{eqnarray*}
Here, after applying a set of positive operator-valued measurements (POVMs) $\mathcal{M}$, the state $\sigma_{AB}$ is mapped to a probability distribution $P_W^{\sigma_{AB}}$, that must lie in the set $\mathcal{Q}$ which is defined as the collection of statistics for which the protocol does not abort.
In our protocol, we follow the same idea as \cite{RR08} where the set $\Gamma$ depends on the error rate $\eta$ and is characterized by the following condition which is as follows: For all measurement bases of Alice and Bob as stated in Algorithm \ref{alg1}, and for all vectors of Alice and Bob with respect to their respective bases (denoted as $v_A$ and $v_B$ respectively) such that $v_B \neq v_\ell$ and
\begin{eqnarray*}
    v_A \neq 
    \begin{cases}
        v_1, & \text{if } v_1 = v_\ell, \\
        v_2, & \text{otherwise}.
    \end{cases}
\end{eqnarray*}
then the following inequality
\begin{eqnarray}\label{atmost}
    \left( \bra{v_B} \otimes \bra{v_A} \right) \sigma_{AB} \left( \ket{v_A} \otimes \ket{v_B} \right) \leq \frac{\eta}{2},
\end{eqnarray} holds.

{It is of note that the equality condition mentioned in \cite{RR08}, is relaxed in our context. The reasoning behind it is due to the fact that for the case when we have only the Bell-state that we require, the left hand side of Equation \ref{atmost} drops to $0$. } Now, we present the following theorem.

\begin{theorem}\label{theoremprotocol}
Let $\mathcal{H}_A$ and $\mathcal{H}_B$ be three-dimensional Hilbert spaces, and let $\sigma_{ABE} \in P(\mathcal{H}_A \otimes \mathcal{H}_B \otimes \mathcal{H}_E)$ be the shared density operator of Alice, Bob, and Eve. Suppose $\sigma_{XYE}$ is the resulting state obtained from $\sigma_{ABE}$ by performing orthonormal measurements on the spaces $\mathcal{H}_A$ and $\mathcal{H}_B$. Then
\begin{eqnarray*}
&&H(X|E) - H(X|Y)\\ &\ge& 1 
+ \lambda_{0} \log_3 \lambda_{0} 
+ \lambda_{1} \log_3 \lambda_{1} 
+ \lambda_{2} \log_3 \lambda_{2}\\
&& 
+ \lambda_{3} \log_3 \lambda_{3}  
+ \lambda_{4} \log_3 \lambda_{4} 
+ \lambda_{5} \log_3 \lambda_{5}\\
&& + \lambda_{6} \log_3 \lambda_{6} 
+ \lambda_{7} \log_3 \lambda_{7} 
+ \lambda_{8} \log_3 \lambda_{8} \\
&& 
- (\lambda_{0} + \lambda_{1} + \lambda_{2}) \log_3 (\lambda_{0} + \lambda_{1} + \lambda_{2})\\
&& + (\lambda_{0}^2 + \lambda_{1}^2 + \lambda_{2}^2) \log_3 (\lambda_{0}^2 + \lambda_{1}^2 + \lambda_{2}^2)\\
&& 
- (\lambda_{3} + \lambda_{4} + \lambda_{5}) \log_3 (\lambda_{3} + \lambda_{4} + \lambda_{5}) \\
&& + (\lambda_{3}^2 + \lambda_{4}^2 + \lambda_{5}^2) \log_3 (\lambda_{3}^2 + \lambda_{4}^2 + \lambda_{5}^2)\\
&& 
- (\lambda_{6} + \lambda_{7} + \lambda_{8}) \log_3 (\lambda_{6} + \lambda_{7} + \lambda_{8}) \\
&& - (\lambda_{6}^2 + \lambda_{7}^2 + \lambda_{8}^2) \log_3 (\lambda_{6}^2 + \lambda_{7}^2 + \lambda_{8}^2)
\end{eqnarray*}

where $\lam_i$'s are eigenvalues of $\sigma_{AB}$ i.e. the density operator of the joint system of Alice and Bob with respect to the 3-dimensional Bell basis mentioned in Section \ref{secqtr}.

\end{theorem}
\pf See Appendix \ref{appdx:A1}. $\hfill\square$

It is important to note that the above theorem characterizes the key rate as the minimum of the specified entropy difference. The proof of Theorem \ref{theoremprotocol} is primarily non-constructive, meaning that it establishes the result in principle through existence arguments, rather than by explicitly constructing the solution.

We identify the exact diagonal matrix in \(\Gamma\) that satisfies the required constraints, by adopting an approximation. This approximation matrix construction arises from the assumption that all of the Bell-states except the first one (which is used in in our measurements) is equally probable. This approximate form is computationally tractable where we choose the diagonal matrix as
\begin{eqnarray*}
\mathrm{diag}\left(1 - c\eta, \underbrace{\dfrac{c\eta}{8}, \ldots, \dfrac{c\eta}{8}}_{\text{8 times}}\right),
\end{eqnarray*}
where $c$ is a tunable parameter. This choice of diagonal matrix comes from Theorem \ref{theoremprotocol}. This simplification allows for meaningful analytical progress while preserving the essential structure required for the analysis. 
We find the general qutrit mutual information, which is  
\begin{eqnarray*}
H(X:Y) &=& H(X) + H(Y) - H(X,Y) \\
       &=& H(\sigma_X) + H(\sigma_Y) - H(\sigma_{XY}) \\
       &=& 2 + \left(\log_3{\left(\lambda_{0}^{2} + \lambda_{1}^{2} + \lambda_{2}^{2} \right)} - 1\right) \left(\lambda_{0}^{2} + \lambda_{1}^{2} + \lambda_{2}^{2}\right) \\
       && + \left(\log_3{\left(\lambda_{3}^{2} + \lambda_{4}^{2} + \lambda_{5}^{2} \right)} - 1\right) \left(\lambda_{3}^{2} + \lambda_{4}^{2} + \lambda_{5}^{2}\right) \\
       && + \left(\log_3{\left(\lambda_{6}^{2} + \lambda_{7}^{2} + \lambda_{8}^{2} \right)} - 1\right) \left(\lambda_{6}^{2} + \lambda_{7}^{2} + \lambda_{8}^{2}\right)
\end{eqnarray*}
Particularly for our case, the mutual information is:
\small\begin{eqnarray*}
H(X:Y) &=& \frac{1}{32} \left(
3 c^2 \eta^2 \log_3\left( \frac{c^2 \eta^2}{64} \right) \right.
\\&&- 8 \left(c^2 \eta^2 + 8 (c \eta - 1)^2\right) 
\log_3\left( \frac{c^2 \eta^2}{24} + \frac{(c \eta - 1)^2}{3} \right) \\
&& \left.+ (c^2 \eta^2 + 32 (c \eta - 1)^2) 
\log_3\left( \frac{c^2 \eta^2}{96} + \frac{(c \eta - 1)^2}{3} \right) \right)
\end{eqnarray*}\normalsize

\begin{widetext}

\begin{figure}[H]
\centering    
    \includegraphics[height=10.5 cm,width=12 cm]{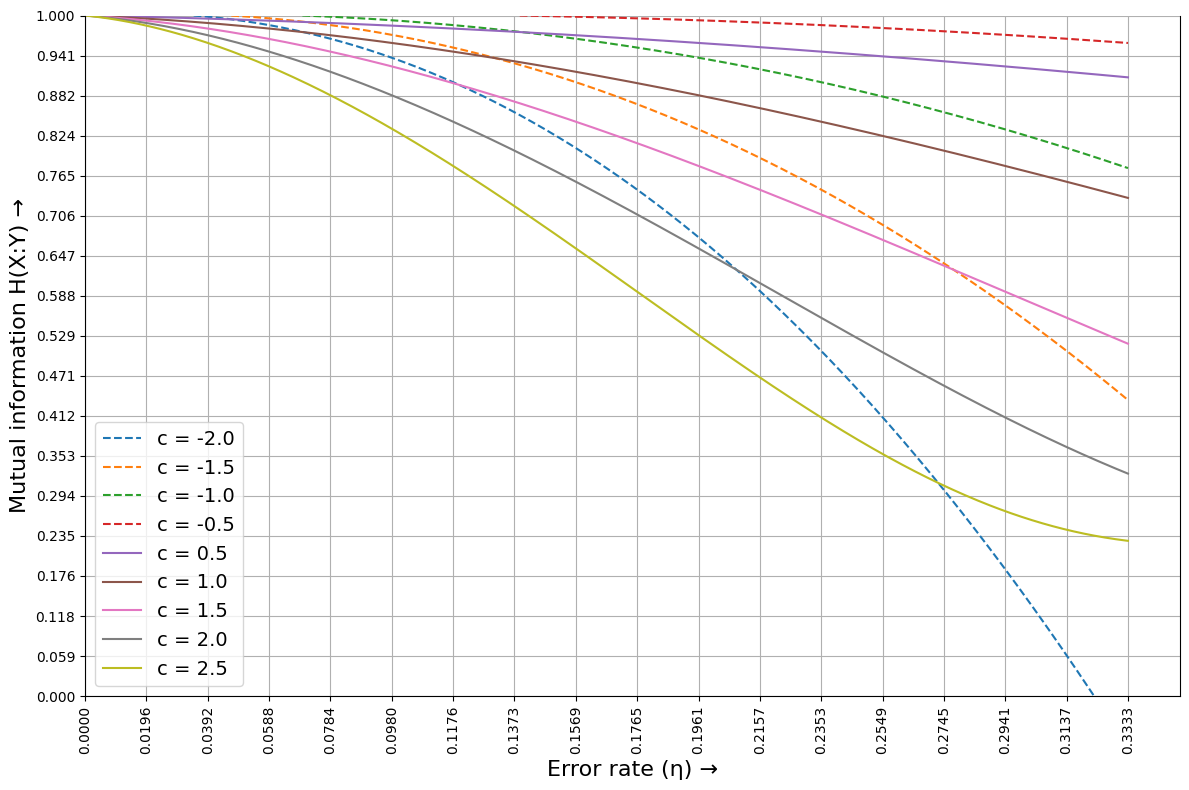}
    \caption{Variation of mutual information $H(X:Y)$ with error rate $\eta$ for different values of the correlation parameter $c$. Mutual information decreases as the error rate increases, with the rate of decline dependent on $c$. Negative values of $c$ (dashed lines) show mutual information greater than $1$, which is contradictory, and are thus discarded, while positive values (solid lines) retain high mutual information at most $1$. The curves correspond to: 
    \textcolor{blue}{dashed blue} for $c = -2.0$, 
    \textcolor{orange}{dashed orange} for $c = -1.5$, 
    \textcolor{green}{dashed green} for $c = -1.0$, 
    \textcolor{red}{dashed red} for $c = -0.5$, 
    \textcolor{violet}{solid violet} for $c = 0.5$, 
    \textcolor{brown}{solid brown} for $c = 1.0$, 
    \textcolor{magenta}{solid magenta} for $c = 1.5$,
    \textcolor{gray}{solid gray} for $c=2.0$, and
    \textcolor{darkolivegreen}{solid olive} for $c=2.5$}.
    \label{fig:enter-label}
\end{figure}
\end{widetext}

We then followed up by analyzing the graph of the mutual information as a function of the error rate, ranging from $0$ to $33.33\%$, and discretized the parameter $c$ from $-2.0$ to $2.5$ in steps of \(0.5\), excluding $c = 0$. We immediately observe that the mutual information values for $c < 0$ can exceed 1, which violates the fundamental bound $H(X:Y) \leq \log_3 3 = 1$ for qutrit systems. In fact, in Figure ~\ref{fig:enter-label}, we plot the mutual information as a function of the error rate (i.e., noise tolerance), using several discretized values of the parameter $c$.

We note that the left-hand side of the inequality simplifies to $\max\left(0,\dfrac13-\dfrac{c\eta}4\right)$, and thus to satisfy the inequality, we note that $c\ge2$ should suffice for $0\le\eta\le\dfrac13$. 

We choose $c = 2$ to maximize the inequality and proceed with the calculations. {Thus, assuming the entangled system is trusted by Alice and Bob to a point where they skip the testing phase of Algorithm \ref{alg:alg1}}, the key rate is: 

\small
\begin{eqnarray}\label{keyrate}
r &\gtrapprox& 
 2 \eta \log{\left(\frac{\eta}{4} \right)} - \frac{3 \eta \log{\left(\frac{3 \eta}{4} \right)}}{2} \\\nonumber&& - \left(2 \eta - 1\right) \log{\left(1 - 2 \eta \right)}\\\nonumber&& + \frac{\left(3 \eta - 2\right) \log{\left(1 - \frac{3 \eta}{2} \right)}}{2} \\\nonumber&&+ \frac{\left(\eta^{2} + 8 \left(2 \eta - 1\right)^{2}\right) \log{\left(\frac{33 \eta^{2}}{8} - 4 \eta + 1 \right)}}{8} + 1
 \end{eqnarray}\normalsize

 This directly follows from Theorem \ref{theoremprotocol}.

\begin{widetext}

\begin{figure}[H]
\centering    
    \includegraphics[height=10.5 cm,width=12.5 cm]{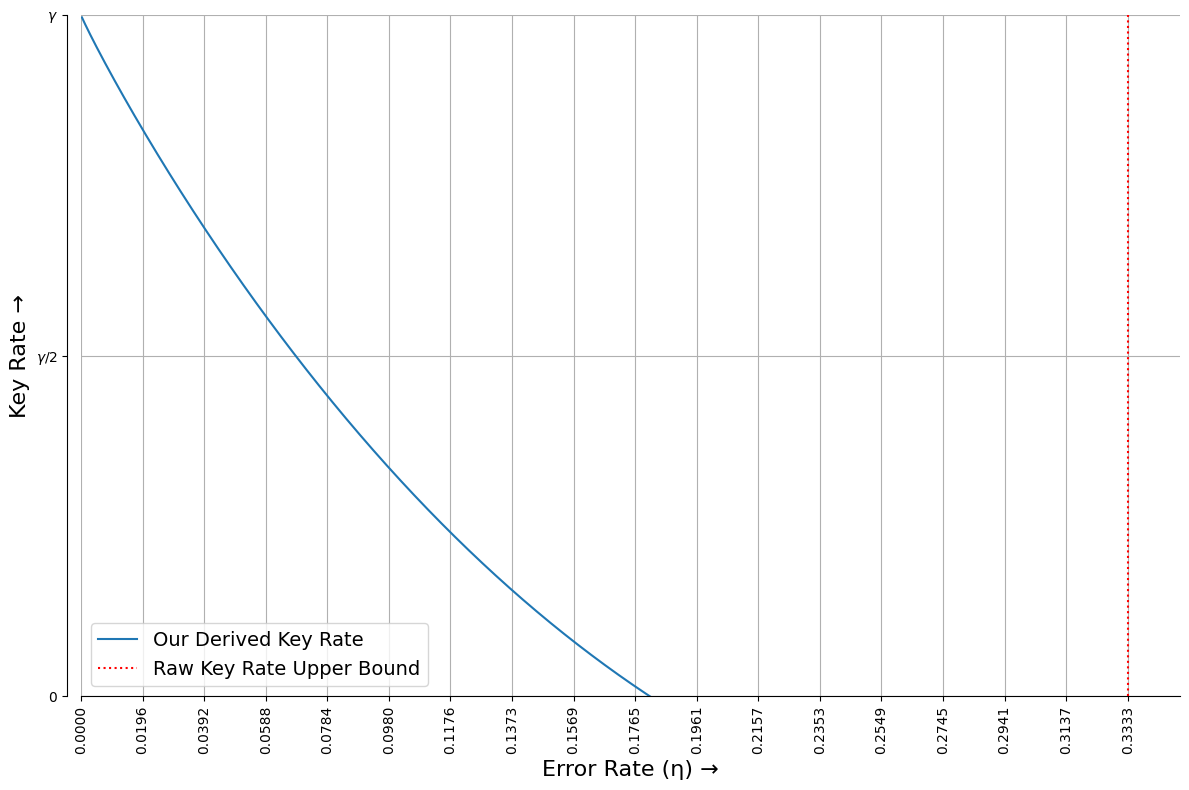}
    \caption{Key rate as a function of error rate $\eta$. The solid \textcolor{blue}{blue} curve represents our derived key rate, which decreases monotonically with increasing error. The vertical \textcolor{red}{red} dashed line indicates the raw key rate upper bound, beyond which secure key generation is not feasible. The plot highlights the trade-off between noise and secure communication rate in the protocol.}\label{fig:enter-label2}
\end{figure}
\end{widetext}

We plot the graph in Figure \ref{fig:enter-label2} for Equation \ref{keyrate} and find that $\text{rate}=0$ when {$\eta\approx18.10610749073431\%$} and beyond.

 This represents a significant advantage, as it implies that the key rate remains constant regardless of the choice of $n$ (Table \ref{tab:notations}). In the ideal scenario where $\eta = 0$, the key rate evaluates to $r = 1$, indicating a 100\% key rate under zero noise.

However, since parameter estimation needs to take place between Alice and Bob, so that they can verify if they receive the state $\dfrac{\ket{00}+\ket{11}+\ket{22}}{\sqrt3}$, a fraction of their outputs with their respective inputs will have to be made public, and thus these cannot influence the key rate. In particular, if $\gamma$ of the total outputs are used to verify the Bell state provided (as stated in the testing phase of Algorithm \ref{alg:alg1}), the new key rate should be $r':=\max\left\{0,\dfrac{\gamma}2 \cdot \inf \sum r(\tilde{\sigma}_{XY})\right\}$, where the infimum is calculated over all possible values of $\tilde{\sig}_{XY}$ such that the corresponding input state ${\sig}_{AB}$ satisfies the Bell inequality as stated in \ref{subsec:3.2}. The factor $\frac12$ is chosen as Bob has two possible inputs, depending on which of the two vectors of Alice he chooses in each of the iterations, whereas Alice's input is fixed. However, we can approximate it, as both the bases chosen by Bob are independent of each other, and thus we get $r'\gtrapprox \max\left\{0,\dfrac{\gamma} 2\cdot 2\cdot r\right\}=\max\left\{0,{\gamma}\cdot r\right\}$. 

Table~\ref{tab3.5} provides a comparison between our protocol and several standard DIQKD protocols, emphasizing key differences such as device assumptions, noise tolerance, the key rate in the zero-noise regime, the dependence on the parameter $n$, and whether a finite-key analysis is included. 

It is also of note that existing DIQKD protocols, thus far, exhibit significantly lower key rates. The highest known key rate is achieved by the BB84 protocol and its variants, where the key rate is given by $R_{\text{BB84}} = 1 - 2h(\eta)$~\cite{PR22}, which is also independent of the parameter $n$. However, as illustrated in Figure \ref{fig:enter-label2}, which plots the key rate $r$ against the noise parameter $\eta$ (error rate), our protocol surpasses the key rate of BB84 (including its variants such as Asymmetric BB84 and BBM92~\cite{BBM92}). Notably, our key rate remains positive until a noise threshold of $18.11\%$, beyond which it drops to zero. This feature is particularly significant for the practical deployment of DIQKD protocols in noisy or near-term intermediate-scale quantum (NISQ) environments. The ability to sustain a comparatively higher key rate in the presence of noise renders our protocol well-suited for implementation on current quantum hardware, where noise is a fundamental limitation.

 \begin{widetext}
  \begin{threeparttable}[t]
\centering
  \begin{tabular}{|c|c|c|c|c|c|c|}
    \hline
    
        \textbf{}&\textbf{Reichardt}&\textbf{Vazirani}&\textbf{Tomamichel}&\textbf{Tomamichel}&\textbf{Zhen}&\textbf{Proposed}\\
        \textbf{}&\textbf{et al. ~\cite{RBFV12}}&\textbf{et al. ~\cite{VV19}}&\textbf{et al.~\cite{TMCNR12}}&\textbf{et al. ~\cite{TMF13}}&\textbf{et al.~\cite{ZM23}}&\textbf{Protocol}\\\hline
         Nature&&& Asymmetric&&&\\
         of Protocol&E91 based &Game based&BB84 based&BBM92 based&Game Based&Game Based\\
         &~\cite{EA91}&~\cite{CHSH69}&~\cite{BB84}&~\cite{BBM92}&~\cite{MD90}&~\cite{KSP90}\\\hline
        Device&&&&&&\\
        Assumptions&None&None&Trusted &Trusted&None&None\\
        &&&Alice*&Alice*&&\\\hline
          Noise&&&&&&\\
          Tolerance&0\%&1.2\%&11\%&1.5\%&3.1\%&\textbf{18.11\%} \\
          &&&&&&\footnotesize{(Highlighted in Figure \ref{fig:enter-label2})}\normalsize\\\hline
          Key rate&&&&&&\\
          (Zero Noise)&0.5\%&2.5\%&100\%&22.8\%&100\%**&100\%** \\
          &&&&&&\\\hline
          Finite key&&&&&&\\
          Analysis&No&No&Yes&Yes&Yes&Yes\\
          &&&&&&\\\hline
        Dependency &&&&&&\\
        on $n$ in &No&Yes&No&No&Yes&No \\key rate&&&&&&\\\hline
      
  \end{tabular}
   \caption{Security Comparison of DIQKD protocols}
    \label{tab3.5}
  \begin{tablenotes}
    \small
    \item * Using results of self-testing ~\cite{LPTR13,TM13}, one can reduce the assumption to memoryless for Alice.

    \item ** The key rate of this protocol depends on the term $\gamma$ where $0<\gamma \le1$. The key rate of $100\%$ is achieved when $\gamma=1$ i.e. the ideal scenario where nothing is discarded in parameter estimation and the noise tolerance is zero.
\end{tablenotes}
\end{threeparttable}


\end{widetext}


\section{Conclusion} \label{conclusion}
 In this paper, we propose a simple ternary device independent quantum key distribution protocol based on the two-party Impossible Colouring pseudo-telepathy game. The simplicity of the protocol arises from the fact that it does not require multiple measurement bases. On the contrary, a single test is sufficient to guarantee correctness, device independence, and raw key security. Moreover, the game is impossible to win classically, which ensures that no classical attack is possible. We demonstrate that the final key security directly follows from the concept of smooth minimum-entropy, which quantifies the randomness in a system and serves as the standard theoretical definition of security in quantum cryptography.

Our protocol employs a qutrit circuit to create $3$-dimensional Bell states, which is advantageous because it requires fewer resources, lower circuit depth, and offers a higher information-carrying capacity compared to traditional qubit-based circuits. This enhances both the efficiency and scalability of the protocol. Furthermore, the use of qutrits contributes to improving the key rate and the robustness of the protocol.

We also show that our proposed protocol exhibits a key rate advantage over established protocols such as BB84 \cite{BB84}, VVQKD \cite{VV14}, and other pseudo-telepathy-based QKD protocols \cite{BMM19,ZM23}. Notably, the key rate is independent of the number of subsystems involved in blockwise measurements, offering scalability benefits. In the future, one could extend this approach to more than two users, leading to a Quantum Conference Key Agreement (QCKA) \cite{MGK20}. Further, the exploration of non-local games, such as the quantum XOR game or non-zero-sum games, could provide alternative foundations for constructing DIQKD protocols that have not yet been thoroughly investigated. We comment that extensive cryptanalytic studies on this protocol would be a valuable direction for future research.

Another important future avenue involves refining the representation of the special set of matching entangled qutrits between two parties, often referred to as the $\Gamma$ set {\cite{RR08}}, which could help optimize our key rate formula and extend the maximum tolerable error bound. Our protocol is inherently one-way and does not rely on noisy preprocessing. However, incorporating advantage distillation and preprocessing could enhance the error threshold beyond the current limit of $18.11\%$. Furthermore, in the future, one may use the Impossible Colouring game and can extend the QKD protocol for more than two users, which is known as Quantum Conference Key Agreement (QCKA)~\cite{MGK20}.

We also mention that if qubits were to be used in the Impossible Colouring game in higher dimensions (i.e., $d = 2^m$ for $m > 1$), one would need to maintain a raw key error rate below $2^{-m}$. This raw key error rate is lower compared to our approach, where we use $d = 3$ and tolerate an error rate up to $\frac1{3}$ (33.33\%), ensuring a more robust security model.

In conclusion, the proposed device-independent quantum key distribution protocol based on the Impossible Colouring pseudo-telepathy game presents a significant advancement in quantum cryptography. Overall, this work lays the groundwork for future developments in secure quantum communication using non-local pseudo-telepathy games.

\section*{Data Availability}
All data and code supporting the findings of this manuscript are available from the authors upon reasonable request.
\section*{Acknowledgments}
 R.S.S. acknowledges funding through MSCA (Grant No. 101126667/ Ramon Llull AIRA). Prof. Avishek Adhikari receives partial support from the DST-FIST Project, funded by the Government of India, under Sanction Order SR/FST/MS-I/2019/41.

\bibliographystyle{unsrt}
\bibliography{paper}

\appendix

\begin{widetext}
\section{Proof of Theorem \ref{thmiv2}}\label{appendix:B}

\pf  In our proposed scheme, the entire protocol operates under the assumption of an ideal, noiseless setting where the communication channel is free from disturbances and the devices used are flawless. However, in practical scenarios, both the channel and the devices are inherently subject to imperfections and noise. Moreover, any adversarial interference is constrained by the laws of quantum physics and would manifest itself in the observed measurement statistics. In the testing phase of our protocol, a noise tolerance parameter $\eta$ is defined to account for these deviations. Accordingly, each raw key bit generated by Alice and Bob may differ with probability at most $\eta$, i.e., for the $i$-th bit of the raw key, we have $\Pr(K_A^i \ne K_B^i) \leq \eta$. This implies that the probability of agreement on each bit satisfies $\Pr(K_A^i = K_B^i) \geq 1 - \eta$.\\

Therefore, if Alice and Bob share a $k$-bit raw key, the probability that their entire raw keys are identical is at least $(1 - \eta)^k$, that is, $\Pr(K_A = K_B) \geq (1 - \eta)^k$. Consequently, the probability that their keys differ is bounded by $\Pr(K_A \ne K_B) < 1 - (1 - \eta)^k$. Setting $\epsilon_{\text{correct}} = 1 - (1 - \eta)^k$, we conclude that the proposed scheme is $\epsilon_{\text{correct}}$-correct, as required. \hfill$\square$

\section{Proof of Theorem \ref{Thd}}\label{appendix:C}
\pf According to the rule of game, both the players start the game with the entangled state $\ket{\psi}:=\displaystyle\frac{1}{\sqrt{d}}\sum_{i=0}^{d-1}\ket{j}\ket{j}$. Alice and Bob will measure a random vector $v_a$ and $v_b$ respectively using their measurement basis and will win the game when $v_a=v_b$. Now, \begin{align*}
       \Pr(A=v_a,B=v_b)&=\left|\bra{\psi}(\ket{v_a}\ket{v_b})\right|^2\\&=\left|\frac{1}{\sqrt{d}}\sum\limits_{j=0}^{d-1}\bra{j}\braket{j|v_a}\ket{v_b}\right|^2\\
    &= \left|\frac{1}{\sqrt{d}}\sum\limits_{j=0}^{d-1}\braket{j|v_a}\braket{j|v_b}\right|^2\\
      &= \left|\frac{1}{\sqrt{d}}\sum\limits_{j=0}^{d-1}\braket{v_a|j}\braket{j|v_b}\right|^2\\
    &=\left|\frac{1}{\sqrt{d}}\braket{v_a|v_b}\right|^2 \\
    &=\begin{cases}
        \dfrac{1}{d} & \text{ if }  v_a=v_b\\
        0 & \text{ otherwise }
    \end{cases}
    \end{align*}
    The winning strategy is thus as follows: Bob measures one of the vectors that is common with Alice, announces the outcome, and Alice subsequently performs her own measurement. Since the measurement basis is orthonormal, the probability that Bob obtains any particular outcome is uniformly distributed among the $d$ basis vectors. The success event can therefore be characterized using the following indicator function defined as \label{indicator}{$$\mathbbm{1}_X:=\begin{cases}
        1 & \text{if }X\text{ is true},\\
        0 & \text{otherwise}
    \end{cases}$$} the probability of Alice getting a vector $v_a$, given Bob has measured $v_b$ is thus: 
    \begin{align*}   \Pr(A=v_a|B=v_b)&=\frac{Pr(A=v_a,B=v_b)}{Pr(B=v_b)}\\&=\dfrac{\dfrac{1}{d}\cdot\mathbbm{1}_{(v_a=v_b)}}{\dfrac1{d}}
    =\mathbbm{1}_{(v_a=v_b)}
    \end{align*} 

Therefore, given the structure of this pseudo-telepathy game, we conclude that, upon a successful round, Alice measures the same vector as Bob with certainty. $\hfill \square$

\section{Proof of Theorem \ref{iv7}}\label{appendix:D}
\pf In the proposed scheme, Alice and Bob each input three classical vectors into their respective black-box devices, and in the measurement phase, each outputs a single raw key bit. Accordingly, both the real system ($S_{\mathrm{real}}$) and the ideal system ($S_{\mathrm{ideal}}$) take three classical vectors as input and produce one classical bit as output.

We assume that the random variables $\mathrm{\texttt{X}_{\mathrm{ideal}}}$ and $\mathrm{\texttt{X}_{\mathrm{real}}}$ represent the outcomes of the ideal and real systems, respectively, when calculating the closeness measure in terms of statistical distance. The two random variables are defined over the set $S = \{0,1\}$. The statistical distance between these two random variables will be as follows.
   \begin{eqnarray*}
      &&\frac{1}{2}\sum\limits_{x\in S}|Pr(\mathrm{\texttt{X}_{ideal}}=x)-\Pr(\mathrm{\texttt{X}_{real}}=x)|\\&=&\frac{1}{2}(|Pr(\mathrm{\texttt{X}_{ideal}}=0)-Pr(\mathrm{\texttt{X}_{real}}=0)|\\&&+|Pr(\mathrm{\texttt{X}_{ideal}}=1)-Pr(\mathrm{\texttt{X}_{real}}=1)|)\\&=&\frac{1}{2}(|\frac{1}{2}-(\frac{1}{2}-\eta)|+|\frac{1}{2}-(\frac{1}{2}-\eta)|)\\&=& \eta 
\end{eqnarray*} This concludes the proof. $\hfill\square$
\section{Proof of Theorem \ref{theoremprotocol}}\label{appdx:A1}

 \pf The proof follows from the discussion in Section \ref{IV.d}., Equation \ref{atmost}. As we know, in our protocol, the set $\Gamma$ depends on the error rate $\eta$ and is characterized by the condition
\begin{eqnarray*}
    \left( \bra{v_B} \otimes \bra{v_A} \right) \sigma_{AB} \left( \ket{v_A} \otimes \ket{v_B} \right) \leq \frac{\eta}{2},
\end{eqnarray*}
for all $v_A$ and $v_B$ denote the measurement bases of Alice and Bob, respectively, such that $v_B \neq v_\ell$ and
\begin{eqnarray*}
    v_A \neq 
    \begin{cases}
        v_1, & \text{if } v_1 = v_\ell, \\
        v_2, & \text{otherwise}.
    \end{cases}
\end{eqnarray*}

We now define a completely positive map (CPM)  \cite{RR08} $\mathcal{D}$ as
\begin{eqnarray*}
    \mathcal{D}(\sigma_{AB}) := \frac{1}{9} \sum_{\tau \in GP} \tau^{\otimes 2} \sigma_{AB} \tau^{\otimes 2},
\end{eqnarray*}
where $GP$ denotes Sylvester's generalized Pauli matrices \cite{Apple05}:
\begin{eqnarray*}
    GP = \left\{ \left( \Sigma_1 \right)^k \left( \Sigma_2 \right)^j \right\}_{k,j=1}^3,
\end{eqnarray*}
with
\begin{eqnarray*}
    \Sigma_1 :=
    \left[
        \begin{array}{ccc}
            0 & 0 & 1 \\
            1 & 0 & 0 \\
            0 & 1 & 0
        \end{array}
    \right], \quad
    \Sigma_2 :=
    \left[
        \begin{array}{ccc}
            1 & 0 & 0 \\
            0 & \omega & 0 \\
            0 & 0 & \omega^2
        \end{array}
    \right], \quad \text{where } 1 + \omega + \omega^2 = 0.
\end{eqnarray*}

Let $\tilde{\sigma}_{AB} := \mathcal{D}(\sigma_{AB})$, and let $\tilde{\sigma}_{ABE}$ be the arbitrary purification of $\tilde{\sigma}_{AB}$, which includes interference from Eve. We define

\begin{eqnarray*}
    \tilde{\sigma}_{XYE} = \left( \mathcal{E}^\text{Meas}_{XY \leftarrow AB} \otimes {I}_{|E|} \right) \tilde{\sigma}_{ABE}.
\end{eqnarray*} 

where $\mathcal{E}^\text{Meas}_{XY \leftarrow AB} : \mathcal{H}_A \otimes \mathcal{H}_B \to \mathcal{H}_X \otimes \mathcal{H}_Y$ denote the measurement map, where, following the measurement in a chosen basis by Alice and Bob, the Hilbert spaces $\mathcal{H}_A$ and $\mathcal{H}_B$ are transformed into the Hilbert spaces $\mathcal{H}_X$ and $\mathcal{H}_Y$, respectively.\\

It is trivial to see that

\begin{eqnarray*}
    \tilde{\sigma}_{AB} = \sum_{i=0}^8 \lambda_i \ket{\varphi_i} \bra{\varphi_i},
\end{eqnarray*}
where $\left\{ \ket{\varphi_i} \right\}_{i=0}^8$ are the 3-dimensional Bell basis (mentioned in Section \ref{secqtr}), denoted as:

\begin{align*}
    \ket{\varphi_0} &= \frac{\ket{00} + \ket{11} + \ket{22}}{\sqrt{3}}, \\
    \ket{\varphi_1} &= \frac{\ket{00} + \omega \ket{11} + \omega^2 \ket{22}}{\sqrt{3}}, \\
    \ket{\varphi_2} &= \frac{\ket{00} + \omega^2 \ket{11} + \omega \ket{22}}{\sqrt{3}}, \\
    \ket{\varphi_3} &= \frac{\ket{01} + \ket{12} + \ket{20}}{\sqrt{3}}, \\
    \ket{\varphi_4} &= \frac{\ket{01} + \omega \ket{12} + \omega^2 \ket{20}}{\sqrt{3}}, \\
    \ket{\varphi_5} &= \frac{\ket{01} + \omega^2 \ket{12} + \omega \ket{20}}{\sqrt{3}}, \\
    \ket{\varphi_6} &= \frac{\ket{02} + \ket{10} + \ket{21}}{\sqrt{3}}, \\
    \ket{\varphi_7} &= \frac{\ket{02} + \omega \ket{10} + \omega^2 \ket{21}}{\sqrt{3}}, \\
    \ket{\varphi_8} &= \frac{\ket{02} + \omega^2 \ket{10} + \omega \ket{21}}{\sqrt{3}}.
\end{align*}

In other words, $\tilde{\sigma}_{AB}$ is a diagonal matrix with respect to the Bell basis. Further, since $\mathcal{D}$ commutes with the measurement operation on $\mathcal{H}_A \otimes \mathcal{H}_B$, it is straightforward to confirm that the conditional entropy $H(X|Y)$ for $\sigma_{XY}$ is bounded above by the corresponding entropy for $\tilde{\sigma}_{XY}$. Similarly, because $\tilde{\sigma}_{ABE}$ is a purification of $\tilde{\sigma}_{AB}$, the conditional entropy $H(X|E)$ for $\sigma_{XE}$ is bounded below by the entropy of $\tilde{\sigma}_{XE}$. Therefore, it is sufficient to establish that the inequality in the lemma holds for the operator $\tilde{\sigma}_{XYE}$, which is derived from the diagonal operator $\tilde{\sigma}_{AB}$.

Let $\ket{e_i}$ be the orthonormal basis Eve uses to interfere with Alice and Bob's communication. Then

\begin{eqnarray*}
    \tilde{\sigma}_{XYE} = \ket{\Psi} \bra{\Psi} \in \mathcal{P}(\mathcal{H}_A \otimes \mathcal{H}_B \otimes \mathcal{H}_E),
\end{eqnarray*}
with $\ket{\Psi}$ defined as:

\begin{eqnarray*}
    \ket{\Psi} = \sum_{i=0}^8 \sqrt{\lambda_i} \ket{\varphi_i}_{AB} \otimes \ket{e_i}_E.
\end{eqnarray*}


We define 
\begin{eqnarray*}
    \ket{f_{x,y}} = \sum_{j=0}^{2} \omega^{jx} \sqrt{\frac{\lambda_{3(y \ominus x) + j}}{3}} \ket{e_{3(y \ominus x) + j}}
\end{eqnarray*}
where $a \ominus b := (a - b) \bmod 3$ and $x, y \in \{0, 1, 2\}$. With $\omega \in \mathbb{C}$ such that $\omega^3 = 1$ and $1 + \omega + \omega^2 = 0$, each of the states of $f$ are as follows:

\begin{enumerate}
    \item $\ket{f_{00}} = \sqrt{\frac{\lambda_0}{3}} \ket{e_0} + \sqrt{\frac{\lambda_1}{3}} \ket{e_1} + \sqrt{\frac{\lambda_2}{3}} \ket{e_2}$
    \item $\ket{f_{01}} = \sqrt{\frac{\lambda_3}{3}} \ket{e_3} + \sqrt{\frac{\lambda_4}{3}} \ket{e_4} + \sqrt{\frac{\lambda_5}{3}} \ket{e_5}$
    \item $\ket{f_{02}} = \sqrt{\frac{\lambda_6}{3}} \ket{e_6} + \sqrt{\frac{\lambda_7}{3}} \ket{e_7} + \sqrt{\frac{\lambda_8}{3}} \ket{e_8}$
    \item $\ket{f_{10}} = \sqrt{\frac{\lambda_6}{3}} \ket{e_6} + \omega \sqrt{\frac{\lambda_7}{3}} \ket{e_7} + \omega^2 \sqrt{\frac{\lambda_8}{3}} \ket{e_8}$
    \item $\ket{f_{11}} = \sqrt{\frac{\lambda_0}{3}} \ket{e_0} + \omega \sqrt{\frac{\lambda_1}{3}} \ket{e_1} + \omega^2 \sqrt{\frac{\lambda_2}{3}} \ket{e_2}$
    \item $\ket{f_{12}} = \sqrt{\frac{\lambda_3}{3}} \ket{e_3} + \omega \sqrt{\frac{\lambda_4}{3}} \ket{e_4} + \omega^2 \sqrt{\frac{\lambda_5}{3}} \ket{e_5}$
    \item $\ket{f_{20}} = \sqrt{\frac{\lambda_3}{3}} \ket{e_3} + \omega^2 \sqrt{\frac{\lambda_4}{3}} \ket{e_4} + \omega \sqrt{\frac{\lambda_5}{3}} \ket{e_5}$
    \item $\ket{f_{21}} = \sqrt{\frac{\lambda_6}{3}} \ket{e_6} + \omega^2 \sqrt{\frac{\lambda_7}{3}} \ket{e_7} + \omega \sqrt{\frac{\lambda_8}{3}} \ket{e_8}$
    \item $\ket{f_{22}} = \sqrt{\frac{\lambda_0}{3}} \ket{e_0} + \omega^2 \sqrt{\frac{\lambda_1}{3}} \ket{e_1} + \omega \sqrt{\frac{\lambda_2}{3}} \ket{e_2}$
\end{enumerate}

We can restate $\ket{\Psi}$ as:
\[
\ket{\Psi} = \sum_{x, y \in \{0, 1, 2\}} \ket{x} \otimes \ket{y} \otimes \ket{f_{x,y}}.
\]

We infer that
\[
\tilde{\sigma}_{X Y E} = \sum_{x, y} \ket{x} \bra{x} \otimes \ket{y} \bra{y} \otimes \ket{f_{x,y}} \bra{f_{x,y}},
\]
because the operator $\tilde{\sigma}_{XY E}$ is obtained from $\tilde{\sigma}_{A B E}$ by orthonormal measurements on $\mathcal{H}_A$ and $\mathcal{H}_B$.

Now, to calculate the key rate, we can calculate the following values: 
i) $H(X|E) \approx H(\tilde{\sigma}_{XE}) - H(\tilde{\sigma}_E)$ and 
ii) $H(X|Y) \approx H(\tilde{\sigma}_{XY}) - H(\tilde{\sigma}_Y)$, where if $\rho$ is a density matrix with eigenvalues $\{a_i\}_{i=1}^n$, then 
\[
H(\rho) = -\text{Tr}(\rho \log_3 \rho) = -\sum a_i \log_3 a_i,
\]
with the base of the logarithm set to $3$, as we are working with qutrits. 

Now, to compute these values, we use partial traces to collapse one of the dimensions. In particular,
\begin{eqnarray*}
    \tilde{\sigma}_{XE} &=& \sum_{y=0}^2 (I_3 \otimes \bra{y} \otimes I_9) \tilde{\sigma}_{XYE} (I_3 \otimes \ket{y} \otimes I_9) \\
    &=& \sum_{y'} (I_3 \otimes \bra{y'} \otimes I_9) \left( \sum_{x, y} \ket{x} \bra{x} \otimes \ket{y} \bra{y} \otimes \ket{f_{x,y}} \bra{f_{x,y}} \right) (I_3 \otimes \ket{y'} \otimes I_9) \\
    &=& \sum_{x, y, y'} \ket{x} \bra{x} \otimes |\braket{y' | y}|^2 \otimes \ket{f_{x,y}} \bra{f_{x,y}} \\
    &=& \sum_{x, y, y'} \ket{x} \bra{x} \otimes \mathbbm{1}_{(y' = y)} \otimes \ket{f_{x,y}} \bra{f_{x,y}} \quad\quad\quad\quad\left[\mathbbm{1}_X\text{ as stated in \ref{indicator} }\right] \\
    &=& \sum_x \ket{x} \bra{x} \otimes \left( \sum_y \ket{f_{x,y}} \bra{f_{x,y}} \right).
\end{eqnarray*}

Similarly, it can be shown that \begin{eqnarray*}
\tilde\sigma_E=\displaystyle\sum_{x,y}\ket{f_{x,y}}\bra{f_{x,y}}
\end{eqnarray*}

Before calculating the entropies as stated above, we first present the matrix form of the corresponding density states. First, we define a sequence of matrices $B_i^{(j)}$ as follows: For $i,j\in \mathbb{Z}_3$ 
\begin{eqnarray*}
B_i^{(j)}:=\dfrac{1}{3}
\left[\begin{array}{ccc}
    \lambda_{3i} & \omega^{2j}\sqrt{\lambda_{3i}\lambda_{3i+1}} & \omega^{j}\sqrt{\lambda_{3i}\lambda_{3i+2}}\\
     \omega^j \sqrt{\lambda_{3i}\lambda_{3i+1}} & \lambda_{3i+1} & \omega^{2j} \sqrt{\lambda_{3i+1}\lambda_{3i+2}}\\
     \omega^{2j}\sqrt{\lambda_{3i}\lambda_{3i+2}} & \omega^{2j} \sqrt{\lambda_{3i+1}\lambda_{3i+2}} & \lambda_{3i+2}
\end{array}\right]
\end{eqnarray*}
Note that all nine matrices have a determinant of zero, implying at least one of their three eigenvalues is \(0\). The sum of the eigenvalues for an index \(i\) is \(\dfrac{\lambda_{3i} + \lambda_{3i+1} + \lambda_{3i+2}}{3}\), and testing it confirms that one of the eigenvalues is \(\dfrac{\lambda_{3i} + \lambda_{3i+1} + \lambda_{3i+2}}{3}\). This leaves us with the matrix \(B_i^{(j)}\) having eigenvalues \(\dfrac{\lambda_{3i} + \lambda_{3i+1} + \lambda_{3i+2}}{3}\) with multiplicity $1$ and $0$ with multiplicity $2$. Now we define a block diagonal matrix 
\begin{eqnarray*}
B^{(j)}:=\left[\begin{array}{c|c|c}
     B^{(j)}_0&0_{3\times3}&0_{3\times3}  \\\hline
     0_{3\times3} & B^{(j)}_1&0_{3\times3}\\\hline
     0_{3\times3}&0_{3\times3}&B^{(j)}_2
\end{array}\right]
\end{eqnarray*}

Then $\tilde{\sigma}_{XE}$ has the matrix form:
\begin{eqnarray*}
\left[\begin{array}{c|c|c}
     B^{(0)}&0_{9\times9}&0_{9\times9}  \\\hline
     0_{9\times9} & B^{(1)}&0_{9\times9}\\\hline
     0_{9\times9}&0_{9\times9}&B^{(2)}
\end{array}\right]
\end{eqnarray*}
Since this matrix is a block diagonal matrix consisting of 9 matrices, the eigenvalues of this block matrix are the list of all eigenvalues of these 9 matrices. So the eigenvalues of \(\tilde{\sigma}_{XE}\) are: \(0\) with multiplicity 18, \(\dfrac{\lambda_{0} + \lambda_{1} + \lambda_{2}}{3}\) with multiplicity $3$, \(\dfrac{\lambda_{3} + \lambda_{4} + \lambda_{5}}{3}\) with multiplicity 3, and \(\dfrac{\lambda_{6} + \lambda_{7} + \lambda_{8}}{3}\) with multiplicity 3. Using the fact that \(0 \log 0 := 0\),
 we have 
\begin{eqnarray*}
    H(\tilde{\sigma}_{XE})&=&-3\left(\dfrac{\lambda_0+\lambda_1+\lambda_2}{3}\log_3\left(\dfrac{\lambda_0+\lambda_1+\lambda_2}{3}\right)
+\dfrac{\lambda_3+\lambda_4+\lambda_5}{3}\log_3\left(\dfrac{\lambda_3+\lambda_4
+\lambda_5}{3}\right)\right.\\
&&\quad\quad\quad\left.+\dfrac{\lambda_6+\lambda_7+\lambda_8}{3}\log_3\left(\dfrac{\lambda_6+\lambda_7+\lambda_8}{3}\right)\right)\\&=&(\lambda_0+\lambda_1+\ldots+\lambda_8)-(\lambda_0+\lambda_1+\lambda_2)\log_3(\lambda_0+\lambda_1+\lambda_2)\\&&\quad\quad  -(\lambda_3+\lambda_4+\lambda_5)\log_3(\lambda_3+\lambda_4+\lambda_5)-(\lambda_6+\lambda_7+\lambda_8)\log_3(\lambda_6+\lambda_7+\lambda_8)\\&=&1-(\lambda_0+\lambda_1+\lambda_2)\log_3(\lambda_0+\lambda_1+\lambda_2)-(\lambda_3+\lambda_4+\lambda_5)\log_3(\lambda_3+\lambda_4+\lambda_5)\\&&\quad\quad-(\lambda_6+\lambda_7+\lambda_8)\log_3(\lambda_6+\lambda_7+\lambda_8) 
\end{eqnarray*}

Similarly, $\tilde\sigma_E$ has the diagonal matrix form of a diagonal matrix: $\text{diag}(\lambda_0,\lambda_1,\ldots, \lambda_8)$. So $H(\tilde\sigma_E)=-\sum_{i=0}^8\lambda_i\log_3(\lambda_i)$

At last we come to the analysis of $H(X|Y)=H(\tilde{\sigma}_{XY})-H(\tilde{\sigma}_Y)$. $\tilde{\sigma}_{XY}$ has the form

\begin{eqnarray*}
\frac{1}{3} 
\scalebox{0.7}{$
\left[
\begin{array}{ccccccccc}
    \lambda_0^2+\lambda_1^2+\lambda_2^2 & 0 & 0 & 0 & 0 & 0 & 0 & 0 & 0 \\
    0 & \lambda_3^2+\lambda_4^2+\lambda_5^2 & 0 & 0 & 0 & 0 & 0 & 0 & 0 \\
    0 & 0 & \lambda_6^2+\lambda_7^2+\lambda_8^2 & 0 & 0 & 0 & 0 & 0 & 0 \\
    0 & 0 & 0 & \lambda_6^2+\lambda_7^2+\lambda_8^2 & 0 & 0 & 0 & 0 & 0 \\
    0 & 0 & 0 & 0 & \lambda_0^2+\lambda_1^2+\lambda_2^2 & 0 & 0 & 0 & 0 \\
    0 & 0 & 0 & 0 & 0 & \lambda_3^2+\lambda_4^2+\lambda_5^2 & 0 & 0 & 0 \\
    0 & 0 & 0 & 0 & 0 & 0 & \lambda_3^2+\lambda_4^2+\lambda_5^2 & 0 & 0 \\
    0 & 0 & 0 & 0 & 0 & 0 & 0 & \lambda_6^2+\lambda_7^2+\lambda_8^2 & 0 \\
    0 & 0 & 0 & 0 & 0 & 0 & 0 & 0 & \lambda_0^2+\lambda_1^2+\lambda_2^2
\end{array}
\right]
$}
\end{eqnarray*}

and $\tilde{\sigma}_Y$ has the form 
\begin{eqnarray*}
\frac13\left[\begin{array}{ccc}
     \lambda_0^2+\lambda_1^2+\ldots+\lambda_8^2&  \\
     & \lambda_0^2+\lambda_1^2+\ldots+\lambda_8^2\\
     &&\lambda_0^2+\lambda_1^2+\ldots+\lambda_8^2
\end{array}\right]=\frac13\cdot I
\end{eqnarray*}

So $\tilde{\sigma}_{XY}$ has 3 distinct eigenvalues $\dfrac{\lambda_0^2+\lambda_1^2+\lambda_2^2}3$, $\dfrac{\lambda_3^2+\lambda_4^2+\lambda_5^2}3$ and $\dfrac{\lambda_6^2+\lambda_7^2+\lambda_8^2}3$, each with multiplicity $3$, while $\tilde{\sigma}_{Y}$ has a single eigenvalue $\dfrac13$ with multiplicity 3. So 
\begin{eqnarray*}
    H(\tilde{\sigma}_{XY})&=&-3\left(\dfrac{\lambda_0^2+\lambda_1^2+\lambda_2^2}{3}\log_3\left(\dfrac{\lambda_0^2+\lambda_1^2+\lambda_2^2}{3}\right)
+\dfrac{\lambda_3^2+\lambda_4^2+\lambda_5^2}{3}\log_3\left(\dfrac{\lambda_3^2+\lambda_4^2
+\lambda_5^2}{3}\right)\right.\\
&&\quad\quad\quad\left.+\dfrac{\lambda_6^2+\lambda_7^2+\lambda_8^2}{3}\log_3\left(\dfrac{\lambda_6^2+\lambda_7^2+\lambda_8^2}{3}\right)\right)\\&=&(\lambda_0^2+\lambda_1^2+\ldots+\lambda_8^2)-(\lambda_0^2+\lambda_1^2+\lambda_2^2)\log_3(\lambda_0^2+\lambda_1^2+\lambda_2^2)\\&&\quad\quad  -(\lambda_3^2+\lambda_4^2+\lambda_5^2)\log_3(\lambda_3^2+\lambda_4^2+\lambda_5^2)-(\lambda_6^2+\lambda_7^2+\lambda_8^2)\log_3(\lambda_6^2+\lambda_7^2+\lambda_8^2)\\&=&1-(\lambda_0^2+\lambda_1^2+\lambda_2^2)\log_3(\lambda_0^2+\lambda_1^2+\lambda_2^2)-(\lambda_3^2+\lambda_4^2+\lambda_5^2)\log_3(\lambda_3^2+\lambda_4^2+\lambda_5^2)\\&&\quad\quad-(\lambda_6^2+\lambda_7^2+\lambda_8^2)\log_3(\lambda_6^2+\lambda_7^2+\lambda_8^2)
\end{eqnarray*}
and 
\begin{eqnarray*}
H(\tilde{\sigma}_Y)=-3(\frac13\log_3\frac13)=1
\end{eqnarray*}

So in the end, we get, 
\begin{eqnarray*}
    \text{Key rate}&\geq&H(X|E)-H(X|Y)=H(\tilde{\sigma}_{XE})-H(\tilde{\sigma}_E)-H(\tilde{\sigma}_{XY})+H(\tilde{\sigma}_Y)\\&=&1-(\lambda_0+\lambda_1+\lambda_2)\log_3(\lambda_0+\lambda_1+\lambda_2)-(\lambda_3+\lambda_4+\lambda_5)\log_3(\lambda_3+\lambda_4+\lambda_5)\\&&\quad\quad-(\lambda_6+\lambda_7+\lambda_8)\log_3(\lambda_6+\lambda_7+\lambda_8)++(\lambda_0^2+\lambda_1^2+\lambda_2^2)\log_3(\lambda_0^2+\lambda_1^2+\lambda_2^2)\\&&\quad\quad+(\lambda_3^2+\lambda_4^2+\lambda_5^2)\log_3(\lambda_3^2+\lambda_4^2+\lambda_5^2)-(\lambda_6^2+\lambda_7^2+\lambda_8^2)\log_3(\lambda_6^2+\lambda_7^2+\lambda_8^2)\\&&\quad\quad+\sum_{i=0}^8\lambda_i\log_3(\lambda_i)
\end{eqnarray*}

The rest of the proof follows immediately. $\hfill\square$

\end{widetext}





\end{document}